\documentclass[aps,pre,noamsfonts,twocolumn,groupedaddress,longbibliography,nofootinbib,showkeys,showpacs]{revtex4-1}

\usepackage{latexsym}
\usepackage{amsfonts}
\usepackage{graphicx}
\usepackage{mathptmx}      
\usepackage{bm}
\usepackage{enumerate}
\usepackage{subfigure}
\usepackage{multirow}

\begin{document}


\title{Quantum annealing with anneal path control:\\ application to 2-SAT problems with known energy landscapes}

\author{Ting-Jui Hsu}
\affiliation{Institute for Advanced Simulation, J\"ulich Supercomputing Centre,\\
Forschungszentrum J\"ulich, D-52425 J\"ulich, Germany}
\author{Fengping Jin}
\affiliation{Institute for Advanced Simulation, J\"ulich Supercomputing Centre,\\
Forschungszentrum J\"ulich, D-52425 J\"ulich, Germany}
\author{Christian Seidel}
\affiliation{Volkswagen Data:Lab, Munich, Germany}
\author{Florian Neukart}
\affiliation{Volkswagen Group of America, San Francisco, USA}
\author{Hans De Raedt}
\affiliation{Zernike Institute for Advanced Materials, University of Groningen, \\
Nijenborgh 4, NL-9747AG Groningen, The Netherlands}
\author{Kristel Michielsen}
\thanks{Corresponding author}
\email{k.michielsen@fz-juelich.de}
\affiliation{Institute for Advanced Simulation, J\"ulich Supercomputing Centre,\\
Forschungszentrum J\"ulich, D-52425 J\"ulich, Germany}
\affiliation{RWTH Aachen University, D-52056 Aachen, Germany}

\date{\today}

\begin{abstract}

We study the effect of the anneal path control per qubit, a new user control feature offered on the
D-Wave 2000Q quantum annealer, on the performance of quantum annealing for solving optimization
problems by numerically solving the time-dependent Schr{\" o}dinger equation for the time-dependent
Hamiltonian modeling the annealing problems. The anneal path control is thereby modeled as a modified linear
annealing scheme, resulting in an advanced and retarded scheme. The considered optimization problems
are 2-SAT problems with $12$ Boolean variables, a known unique ground state and a highly
degenerate first excited state. We show that adjustment of the anneal path control can result in a
widening of the minimal spectral gap by one or two orders of magnitude and an enhancement of the success probability
of finding the solution of the optimization problem. We scrutinize various iterative methods based on
the spin floppiness, the average spin value, and on the average energy and describe their performance
in boosting the quantum annealing process.

\end{abstract}


\maketitle

\section{Introduction}\label{INTRODUCTION}

Quantum annealing is a quantum version of classical simulated annealing~\cite{KIRK83}, using quantum fluctuations
instead of thermal fluctuations, to explore the energy landscape of an optimization problem~\cite{KADO98}.
This approach has received enormous interest in the last two decades~\cite{FARH01,DAM01, AHAR07, JANS07, BRAV08, AMIN09, BRAV10, NEUH11}
and is regarded as a second model of quantum computing, which is quite distinct to the gate-based model of quantum computing~\cite{NIEL10}.

A recent overview of theoretical work on adiabatic quantum computation (annealing) in closed systems described by
various types of Hamiltonians, thereby discussing, among many other issues, universality and quantum speedup,
is given in Ref.~\cite{ALBA18}.
Because of the ongoing academic discussions about the universality and hypothetical quantum speedup
for different types of Hamiltonians and because in practice, it is not straightforward to build
devices that are described by the desired Hamiltonian, the efforts to implement quantum annealing
in physical systems for the purpose of actual problem solving are rather limited compared
to those for building gate-based quantum information processors.

Since 1999, D-Wave Systems Inc.~\cite{DWAVE} manufactures quantum annealers as integrated circuits of superconducting
qubits that can be described by the Ising model in the transverse field on a Chimera lattice~\cite{JOHN11}.
This type of quantum annealer was designed for solving quadratic unconstrained binary optimization problems.
Since D-Wave Systems installed its first commercial system in 2010, the D-Wave quantum processor doubles in size
almost every two years and each new generation of machines comes with new user control features.
Their current processor has more than $2000$ qubits. This makes the D-Wave 2000Q machine an interesting system
for researchers to explore, test and utilize it for all kinds of optimization and machine learning
problems~\cite{DICK12,PUDE14,BOIX14,HALL15,NOVO16,LANT17,ANDR17,HARR17,NEUK17,AMIN18}.

The latest D-Wave quantum annealer, the D-Wave 2000Q comes with an increased user control over
the anneal path by means of two new features.
The Anneal Offset feature allows a user to advance or delay the annealing path of individual qubits
and the Anneal Pause \& Ramp feature allows a user to first introduce a pause in the annealing process
at a given point in time and of a certain duration and then rapidly quench the transverse field.
The potential of the Anneal Offset feature, which amounts to tuning the transverse field on an individual qubit basis,
has been demonstrated for a 24-qubit system~\cite{LANT17} and for an integer factoring circuit~\cite{ANDR17}.
In Ref.~\cite{ANDR17}, it was shown that it gives a remarkable improvement over baseline performance,
making the computation more than 1000 times faster in some cases. The Anneal Pause \& Ramp feature
has been used in a study of the three-dimensional transverse Ising model~\cite{HARR17}.

How these user control features influence the quantum dynamics of the system and affect the performance
of solving problems is not well-understood. As mentioned in Ref.~\cite{ANDR17}, determining an optimal set of anneal offsets
for a given optimization problem is difficult. In Ref.~\cite{LANT17}, a heuristic algorithm, similar
to the one described in Ref.~\cite{DICK12}, is used to control the anneal offset. The goal of the present work is (1)
to get more insight in the relation between controlling the quantum dynamics of the system by an anneal offset
and its influence on finding the solution of an optimization problem, and
(2) to assess the effectiveness of several simple algorithms used to control the annealing offsets.
For these purposes, we consider specially designed 2-SAT problems
with 12 Boolean variables and perform quantum annealing by solving the time-dependent Schr{\" o}dinger equation (TDSE).
The advantage of this approach is that, on the one hand, for these small problems with a known and unique solution,
we can calculate the lowest levels of the energy spectrum of the system during the annealing process.
This allows us to study changes in the energy spectrum due to a controlled modification of the annealing process.
On the other hand, this approach also makes it possible to examine which influence the modified
energy spectrum has on the chance for finding the solution of the 2-SAT problem.
We perform a comparative study for 2-SAT problems with different ``hardness'' characteristics
to study correlations between the control parameters in the annealing process
and the success rate of finding the solution. Such correlations could be of great help
in solving other more practical optimization problems on D-Wave quantum processors.

The  paper is structured as follows. In Sec.~\ref{sec2SAT}, we introduce the 2-SAT problem and
transform it into an Ising spin Hamiltonian. In Sec.~\ref{secMETHOD}, we introduce the Hamiltonian
describing a quantum annealer and the concepts of spin floppiness, state degeneracy, perturbative anticrossing and their interrelation.
We also describe the method for simulating the quantum annealing processor to solve the 2-SAT problems.
The simulation results are presented in Sec.~\ref{secRESULTS}. Conclusions are given in Sec.~\ref{secCONCLUSION}.

\section{2-SAT problems }\label{sec2SAT}

A 2-SAT instance is formulated as a Boolean expression in the form of a conjunction
(a Boolean AND operation) of clauses, where each clause is a disjunction (a Boolean OR operation) of two literals.
The literals are Boolean variables or their negations. In this paper we consider 2-SAT problems
that are designed to be hard for classical annealing algorithms~\cite{NEUH14} with $j=1, \ldots , 12$ Boolean variables $x_j=0,1$,
$k=1,\ldots,13$ clauses and thus $26$ literals $L_{k,l}$ with $l=1,2$.
The 2-SAT problem is to find a truth assignment to the Boolean variables that makes the formula
$G=(L_{1,1} \vee L_{1,2} )\wedge(L_{2,1}\vee L_{2,2} )\wedge\cdots\wedge(L_{13,1}\vee L_{13,2} )$ true.
If $G=1$ then the 2-SAT instance is satisfiable.

In order to solve a 2-SAT instance by means of quantum annealing on a D-Wave machine
the problem first has to be encoded in the Ising Hamiltonian.
For this purpose the Boolean variables $x_j=0,1$ are transformed in the spin variables $\sigma_j^z=-1,+1$ using $x_j=(1-\sigma_j^z )/2$.
Next, the problem has to be mapped on the Chimera architecture of the D-Wave quantum processor.
Some of the designed hard 2-SAT problems can be directly mapped on this architecture and
others require the usage of so-called logical qubits for the mapping.
In the direct mapping there is a one to-one correspondence between the Ising spins in the model
and the physical qubits of the quantum processor. In the non-direct mapping
one Ising spin is represented by one logical qubit which consists of a group of physical qubits.

2-SAT instances are solvable polynomial in time.
Nevertheless, different solvers might find different 2-SAT instances easy or hard to solve.
We consider 2-SAT instances that, in Ising-spin language, have a known unique ground state
(and corresponding energy) which is separated from a number of first-excited states,
this number increasing exponentially with the number of Boolean variables $N$~\cite{NEUH14}.
For $N=12$, the 2-SAT problems that we consider have minimal spectral gaps
in the energy spectrum of the system during annealing that range between $0.01$GHz and $3$GHz.
For comparison, for D-Wave quantum processors $k_B T$ ranges between $1.5$GHz and $3$GHz,
where $k_B$ denotes Boltzmann's constant and $T$ the finite operating temperature of the D-Wave quantum annealer.
Hence, the considered 2-SAT problems have rather small minimal spectral gaps which makes them hard to solve by means of quantum annealing.

\section{Simulation method}\label{secMETHOD}

The quantum annealing process can be modeled as a dynamical process of a quantum spin-1/2 system and
can therefore be simulated on a conventional digital computer by solving the TDSE
for a given time-dependent Hamiltonian.
We make use of a Suzuki-Trotter product formula algorithm to solve the TDSE~\cite{RAED06}.

\subsection{Hamiltonian including the anneal offset}\label{secHAM}

\begin{figure}
\begin{center}
\includegraphics[width=8cm]{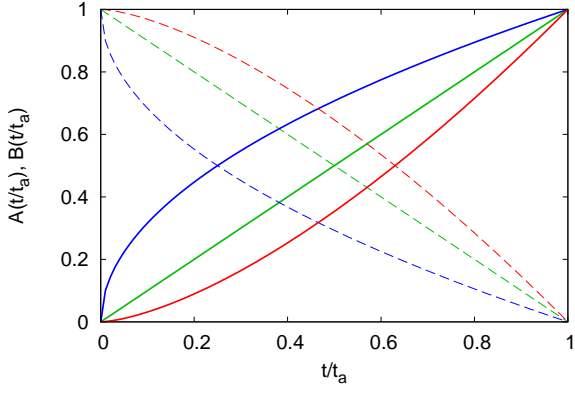}
\caption{%
(Color online) Linear (green), retarded (red) and advanced (blue) annealing schemes for qubit $i$.
The dashed lines indicate $A_i(t/t_a)=1-(t/t_a)^{1+\gamma_i}$ with $\gamma_i=-0.5$, $0$, and $0.5$ (bottom to top).
The solid lines indicate $B_i(t/t_a)=(t/t_a)^{1+\gamma_i}$ with $\gamma_i=-0.5$, $0$, and $0.5$ (top to bottom).}
\label{fig1}
\end{center}
\end{figure}

In general, the quantum annealing Hamiltonian can be written as a time-dependent linear combination of an initial
Hamiltonian $H_I$ and a final Hamiltonian $H_P$ encoding the problem to be solved:
\begin{eqnarray}
H(t) &=& A(t/t_a) H_I + B(t/t_a) H_P, \\
H_I &=& -\sum_{i=1}^N \sigma_i^x, \\
H_P &=& -\sum_{i=1}^N h_i^z \sigma_i^z  - \sum_{i,j=1}^N J_{ij}^z \sigma_i^z\sigma_j^z,
\label{problemH}
\end{eqnarray}
where $t_a$ denotes the total annealing time, $\sigma_i^{x,y,z}$ are Pauli matrices,
$h_i^z$ and $J_{ij}^z$ are real-valued parameters, and $i,j=1,\ldots,N$ label the qubits.
During the quantum annealing process, $A(t/t_a)$ starts from $1$ and slowly decreases to $0$,
and $B(t/t_a)$ starts from $0$ and slowly increases to $1$,
where $A(t/t_a)$ and $B(t/t_a)$ are specified in dimensionless units.
Here we consider
a linear annealing scheme for the quantum annealing process so that $A(t/t_a) = 1 - t/t_a$
and $B(t/t_a) = t/t_a$.
The anneal offset is implemented by modifying $A(t/t_a)$ and $B(t/t_a)$ for each qubit $i$, resulting in the Hamiltonian
\begin{eqnarray}
H(t) &=& -\sum_{i=1}^N A_i(t/t_a)\sigma_i^x-\sum_{i=1}^N B_i(t/t_a) h_i^z \sigma_i^z  \cr
&& - \sum_{i,j=1}^N J_{ij}^z \sqrt{B_i(t/t_a)B_j(t/t_a)} \sigma_i^z\sigma_j^z,
\end{eqnarray}
where $A_i(t/t_a)=1-(t/t_a)^{1+\gamma_i}$ and $B_i(t/t_a)=(t/t_a)^{1+\gamma_i}$.
Examples of linear ($\gamma_i=0$), retarded ($\gamma_i=0.5$), and advanced ($\gamma_i=-0.5$)
annealing schemes are shown in Fig.~\ref{fig1}.

\subsection{Solving the time-dependent Schr{\" o}dinger equation}
The whole system evolves according to the TDSE
\begin{equation}
i\hbar\frac{\partial}{\partial t} \left | \Psi(t)\right > = H(t) \left | \Psi(t)\right >,
\end{equation}
where $\hbar$ is set to unity.
The units used in the simulation are dimensionless
but in Sec.~\ref{secRESULTS} we present the simulation results in units appropriate for the D-Wave machine.
For the time-dependent Hamiltonian $H(t)$, the TDSE is solved by using a small time interval $\tau$ and
\begin{equation}
\left | \Psi(t+\tau)\right > =U(t,\tau)\left| \Psi(t)\right > \approx e^{-i\tau H(t+\tau/2)}\left | \Psi(t)\right >,
\end{equation}
where $H(t+\tau/2)$ is regarded as time-independent, taking its value at time $t+\tau/2$,
and $U(t,\tau) \approx e^{-i\tau H(t+\tau/2)}$ is the time-evolution operator.

For small systems, $U(t,\tau)$ can be calculated by exact diagonalization of the Hamiltonian.
For large systems, when there is not enough memory to store $H$
or $U(t,\tau)$, we use the second-order Suzuki-Trotter product formula~\cite{RAED06}
\begin{eqnarray}
\widetilde{U}_2(t,\tau)&=&
\prod_{j=1}^N e^{i \tau [ A(t/t_a+\tau/2/t_a) \sigma^x_j + B_j(t/t_a+\tau/2/t_a) h^z_j\sigma^z_j]/2 }
\nonumber \\
&&\times
\prod_{j,k=1}^N e^{i \tau [  \sqrt{B_j(t/t_a+\tau/2/t_a)B_k(t/t_a+\tau/2/t_a)} J^z_{j,k}\sigma^z_j\sigma^z_k] }
\nonumber \\
&&\times
\prod_{j=1}^N e^{i \tau [ A(t/t_a+\tau/2/t_a) \sigma^x_j + B_j(t/t_a+\tau/2/t_a) h^z_j\sigma^z_j]/2 },
\label{Utilde}
\end{eqnarray}
to approximate the time-evolution operator $U(t,\tau)$.
This error of this approximation is bounded by $||\widetilde{U}_2(t,\tau)-U(t,\tau)||\leq c_2 \tau^3$,
where $c_2$ is a positive constant~\cite{RAED06}.

\subsection{Floppiness, degeneracy, and perturbative anticrossing}

In an Ising-spin model, a spin is said to be floppy if flipping the spin results in another classical state with the same energy.
Hence, spin floppiness relates to state degeneracy. In Ref.~\cite{LANT17}, an argument, based on first-order perturbation calculations,
is given to relate observed probabilities of qubit floppiness to the energy change of degenerate excited states.
According to a first-order perturbative calculation~\cite{LANT17}, there is a relation between a reduction in excited state energy
and of the minimal spectral gap, and floppiness, namely
\begin{equation}
\Delta'=\frac{1}{2} \sum_{i=1}^{N} A(t/t_a) F_i,
\end{equation}
where $F_i$ is the fraction of degenerate states in which the $i$-th qubit is floppy.

As the performance of quantum annealing is primarily determined by
the minimum spectral gap between the ground state and the first excited state,
we only consider the floppiness of spins in the first excited states of the problem Hamiltonian.
This motivates our choice for considering 2-SAT problems which have a known ground state and a large number
of first-excited states (see section~\ref{sec2SAT}). For these problems, during the annealing process
a perturbative anticrossing could be formed, creating an extremely small minimal spectral gap between
the ground state and the first excited state and causing a slowdown in solving the optimization problem.

A way to prevent the occurrence of such anticrossings is to make use of an anneal offset for individual qubits.
In Ref.~\cite{LANT17} it has been suggested to reduce $\Delta '$ by decreasing $A_i (t/t_a )$ for qubit $i$ for which $F_i$
is high relative to $F_j$ of other qubits $j$. In other words, an advanced anneal offset $A_i (t/t_a)$ should be applied to qubits with a high $F_i$.

\subsection{Simulation procedure}

We use in-house software to solve the TDSE with the non-uniform quantum annealing Hamiltonian described in section~\ref{secHAM}.
In this section, we describe the simulation procedure to solve the 2-SAT problems by a non-uniform quantum annealing process,
thereby mimicking the procedure performed on a D-Wave machine.

First, we compute the parameters $h_i^z$ and $J_{ij}^z$ of the selected 2-SAT problem and fix the annealing offsets $\gamma_i$,
and supply them as input to the software which simulates the annealing process.
The software solves the TDSE for these input parameters and produces a number of output events.
These events consist of strings of $0$s and $1$s, representing the spin states, and the energies corresponding to these spin states.
Besides these events, which are similar to the ones provided by a D-Wave machine, the software also calculates
the probability of finding the ground state (corresponding to the optimal solution) of the problem Hamiltonian
and the average energy $\left < E(t/t_a )\right >$. Finally, we use the output events to compute the single-qubit
average $\overline{\sigma}_i^z$, the final average energy $\overline{E}_f$,
and the floppiness $\mu_i$ of a single qubit $i$.
The latter is given by the ratio of the number of pairs of the degenerate first excited states and the total
number of first excited states.
The single-qubit average $\overline{\sigma}_i^z$ and the final average energy $\overline{E}_f$ based on the generated output events
will be very close to the values calculated directly from the final wave function if the number of output events is large.

The energy spectra of the time-dependent Hamiltonians are calculated by exact diagonalization or by the Lanczos method.

\section{Simulation results}\label{secRESULTS}

The simulation results are presented in two parts.
The first part includes the results obtained by applying an anneal offset to individual qubits.
The second part includes the results obtained by applying the anneal offset to all the qubits in the system.
Systematic iterative methods have been used to dynamically adjust the annealing offsets for each qubit.

\subsection{Annealing offset applied to individual qubits}\label{secOFFONE}

\begin{figure*}
\begin{center}
\includegraphics[width=0.35\hsize]{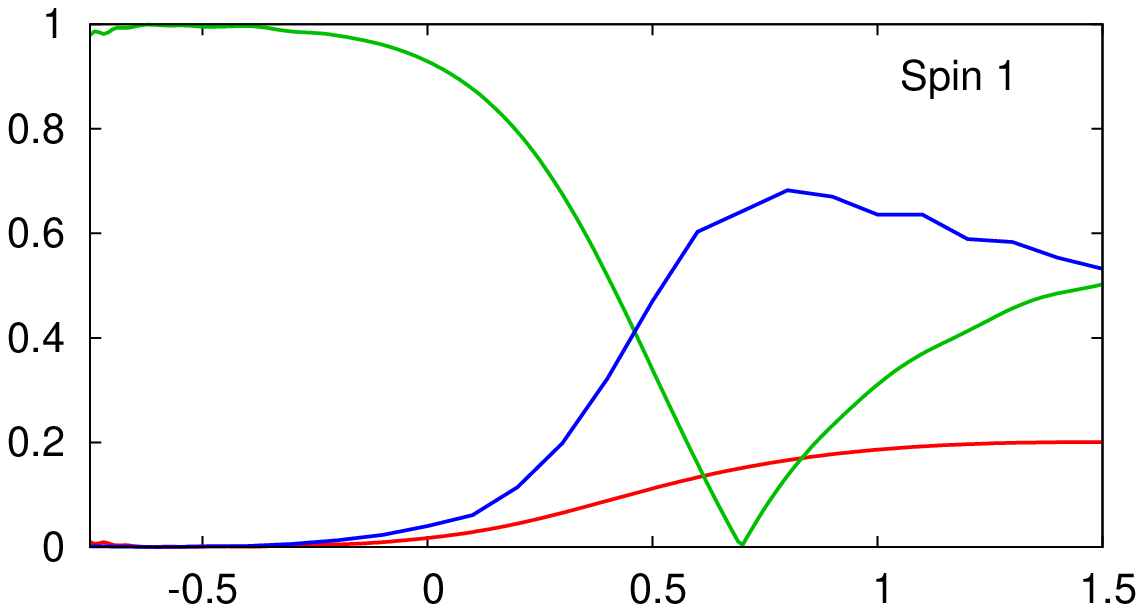}\includegraphics[width=0.35\hsize]{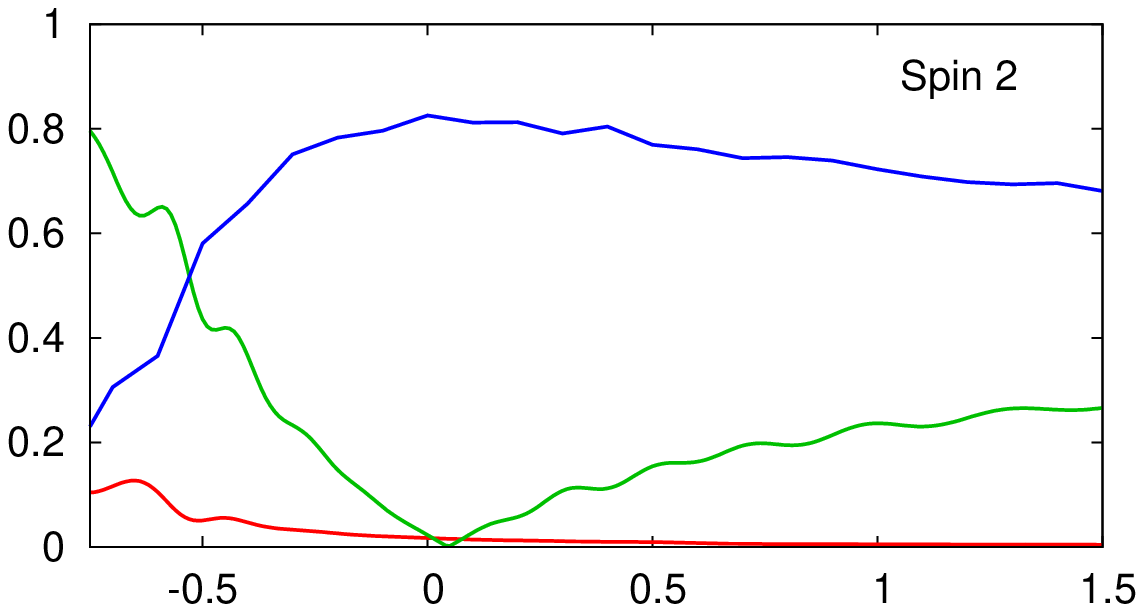}
\includegraphics[width=0.35\hsize]{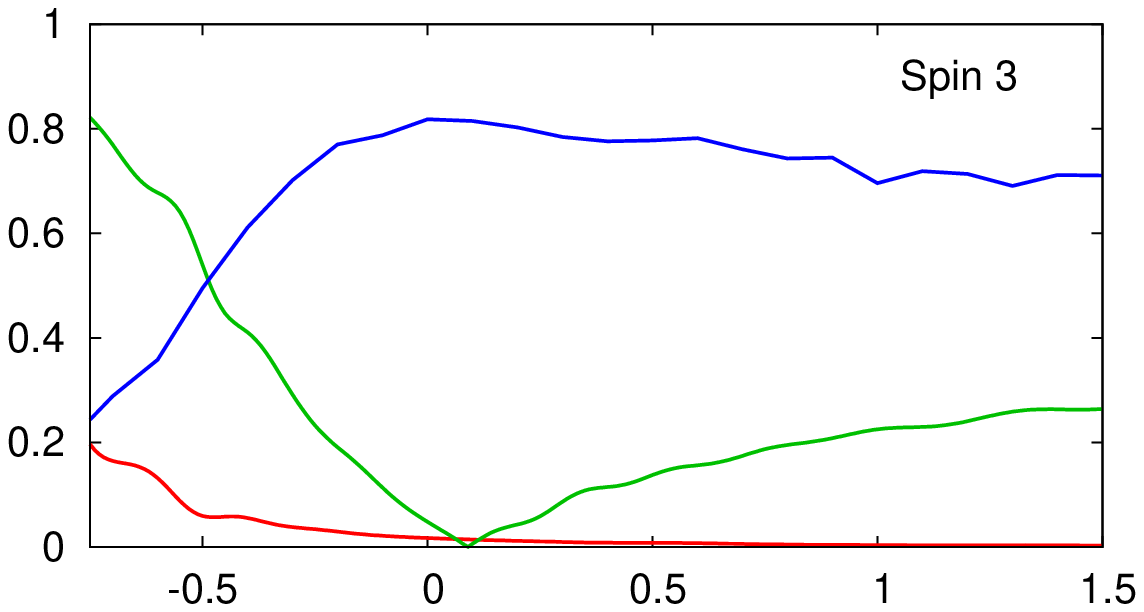}\includegraphics[width=0.35\hsize]{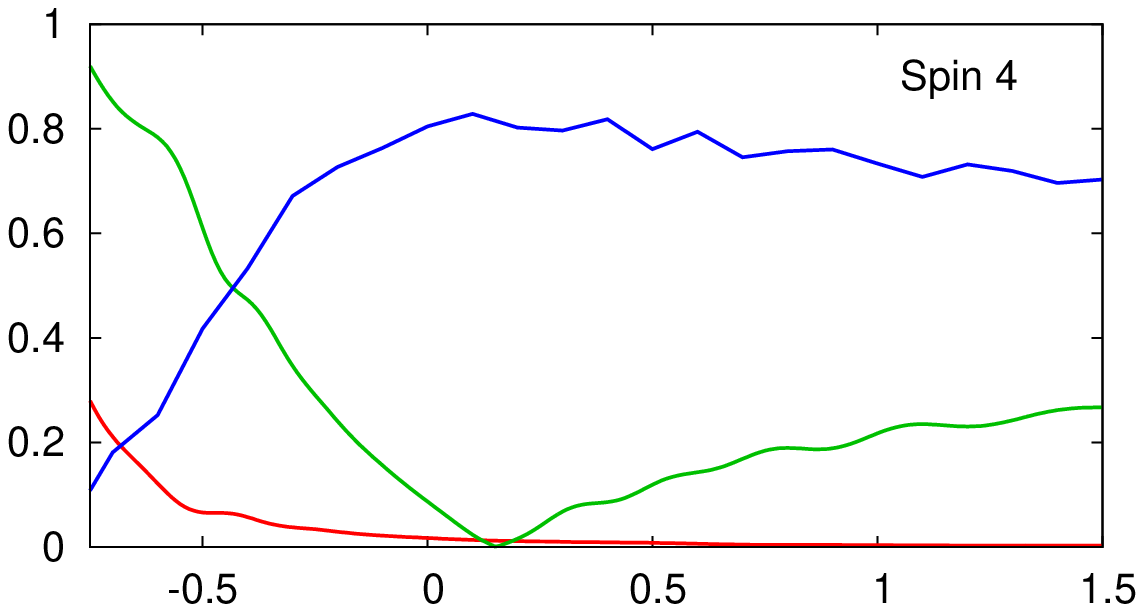}
\includegraphics[width=0.35\hsize]{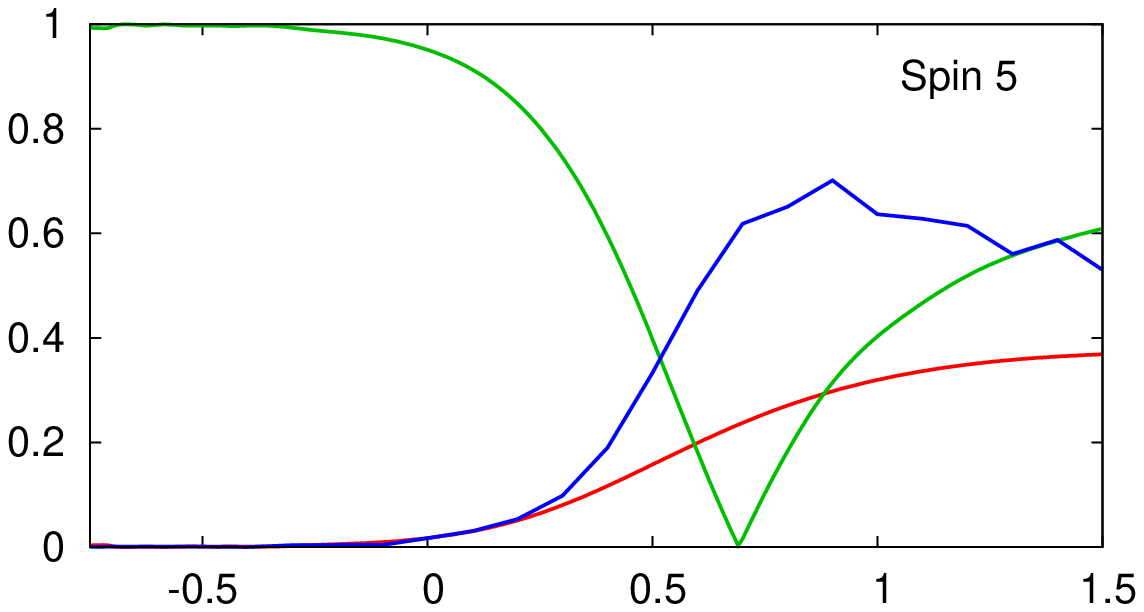}\includegraphics[width=0.35\hsize]{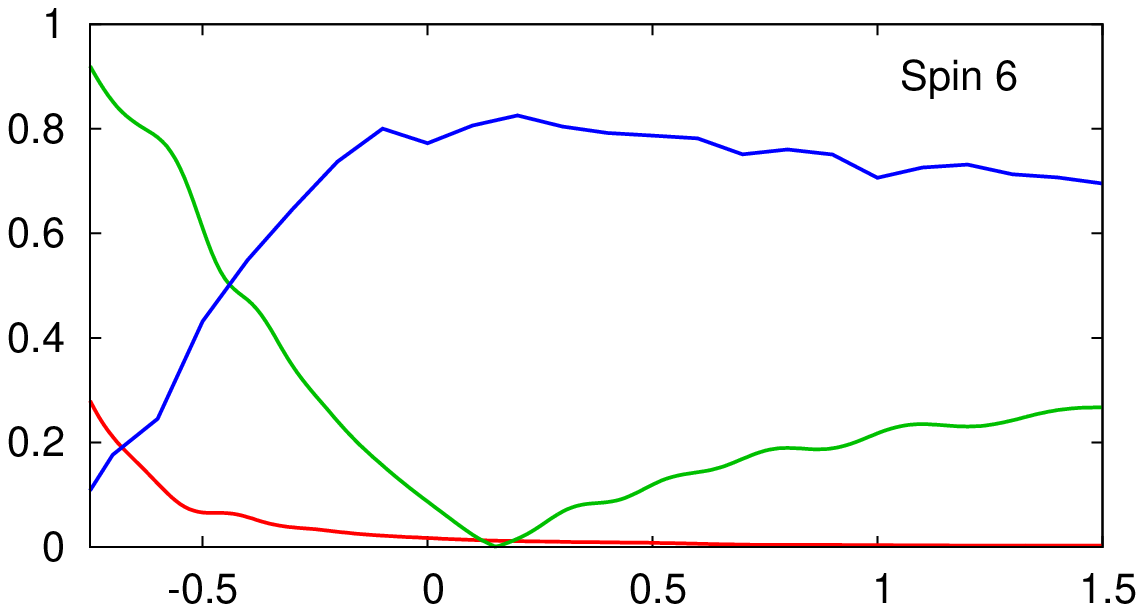}
\includegraphics[width=0.35\hsize]{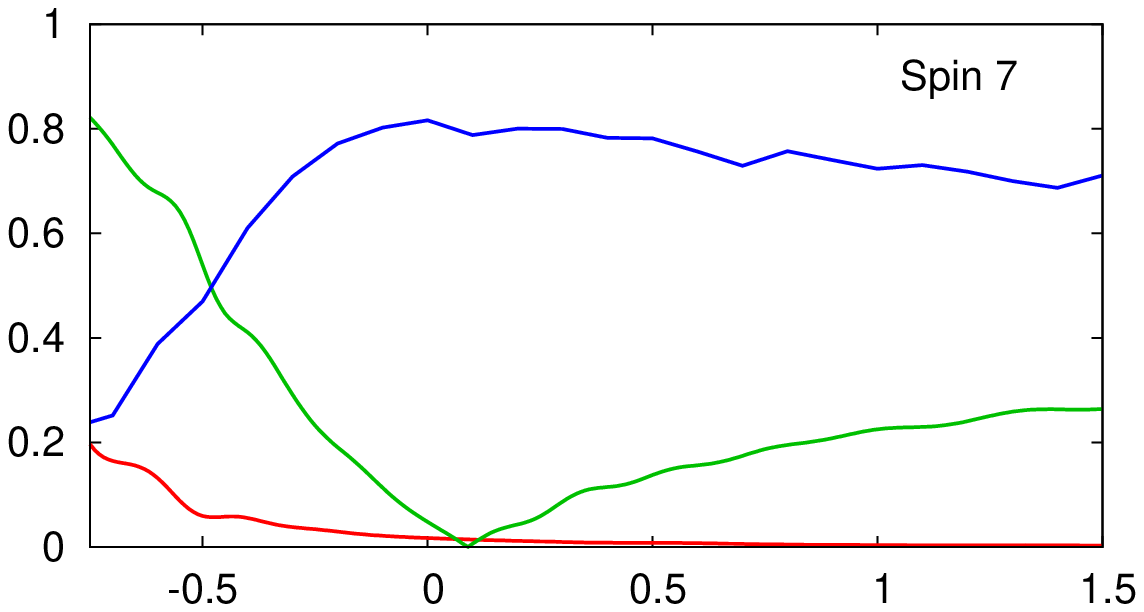}\includegraphics[width=0.35\hsize]{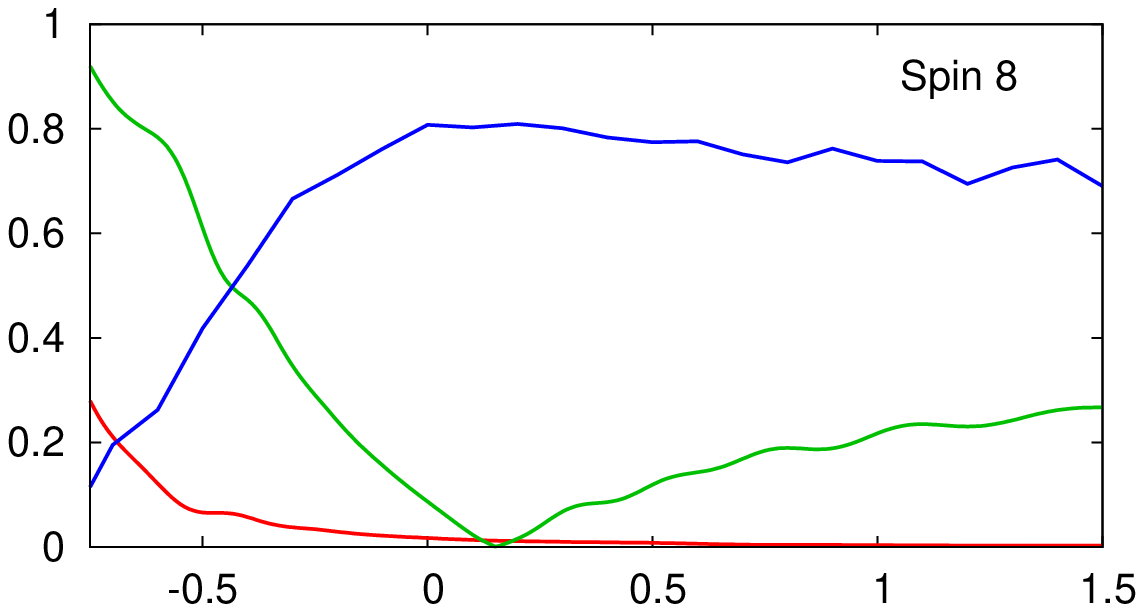}
\includegraphics[width=0.35\hsize]{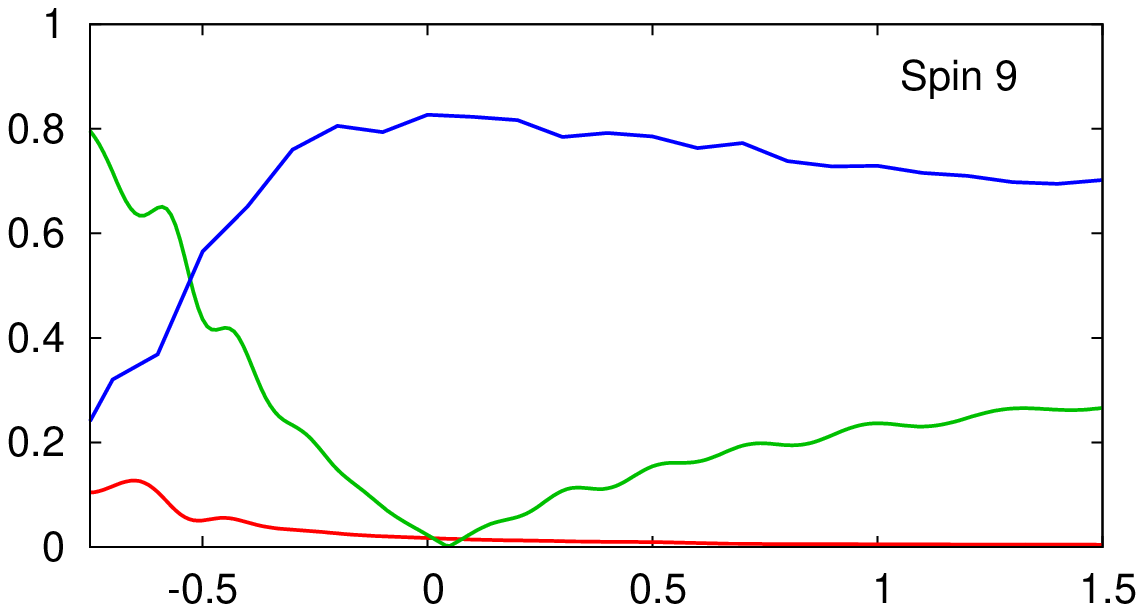}\includegraphics[width=0.35\hsize]{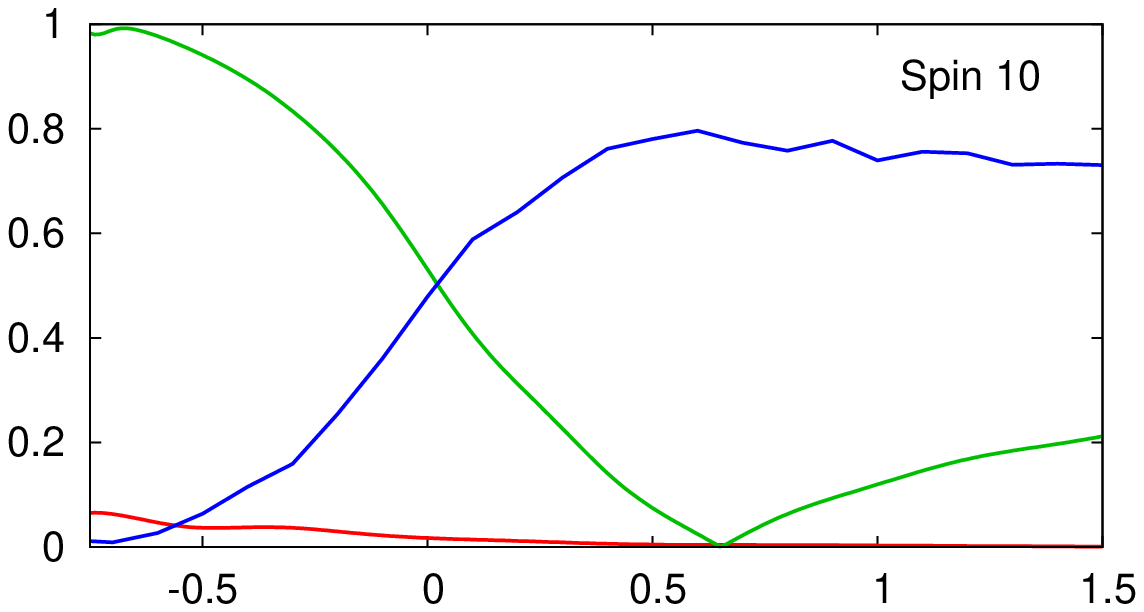}
\includegraphics[width=0.35\hsize]{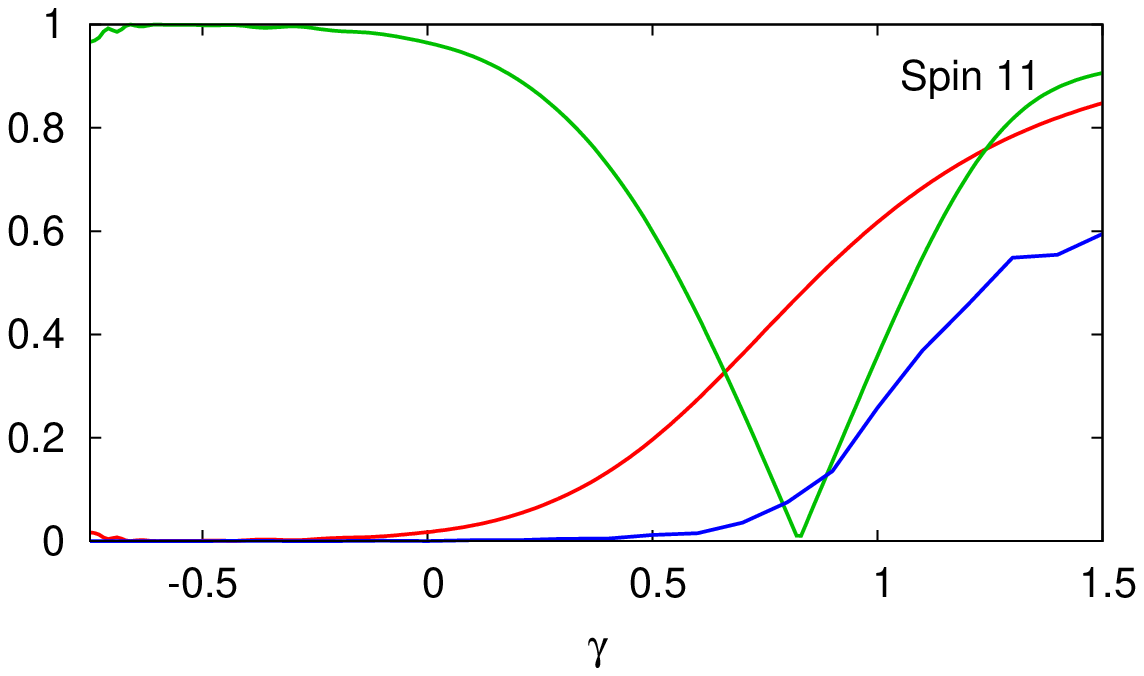}\includegraphics[width=0.35\hsize]{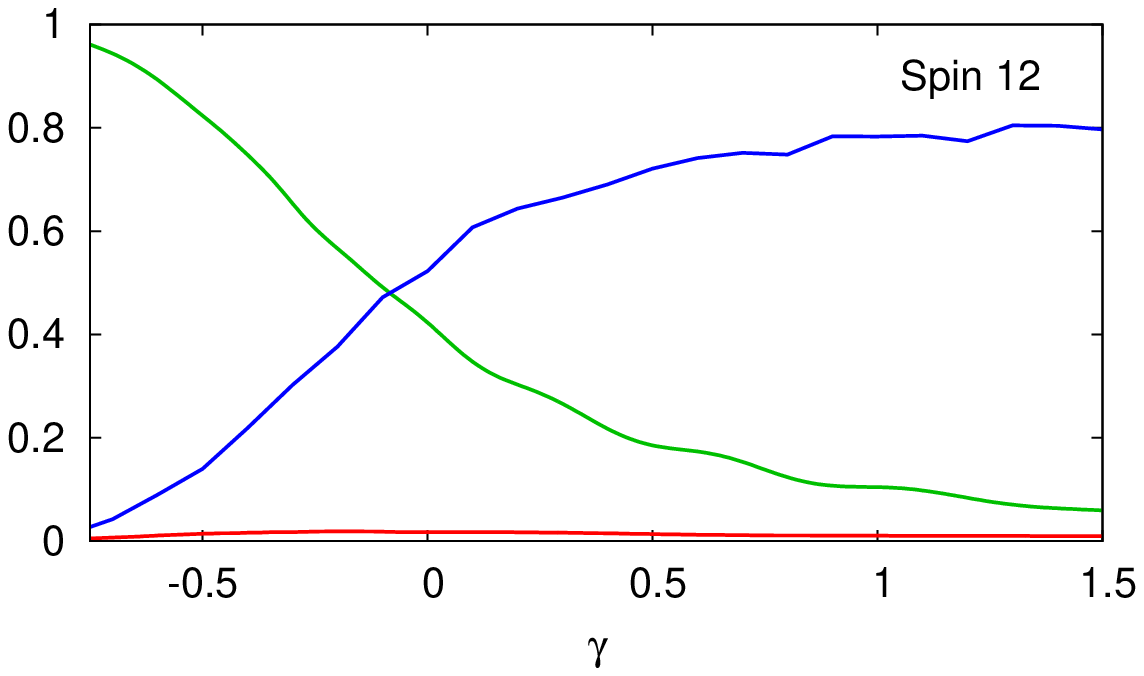}
\caption{%
(Color online) Absolute value of the single-qubit average (green),
floppiness (blue) and the probability for finding the ground state (red)
as a function of the anneal offset $\gamma$ for each spin in the 12-spin 2-SAT problem with number 487.}
\label{fig2}
\end{center}
\end{figure*}

For illustrative purposes, we select one particular problem, problem number $487$ (see the Appendix for the explicit Hamiltonian $H_P$),
of the set of 2-SAT problems with $12$ Boolean variables and $13$ clauses
and having one unique ground state and a large number of first excited states. We adjust the annealing offset $\gamma_i$
for the $i$-th qubit ($i,=1,\ldots,12$) while not touching the others, i.e. $\gamma_{j\neq i}=0$.
The total annealing time is set to $t_a=5$ns.
Figure~\ref{fig2} shows the probability for finding the ground state, the absolute value of the single-qubit average
$|\overline{\sigma}_i^z|$ and
the floppiness as a function of the anneal offset for each qubit.

We observe that altering the linear annealing scheme for each individual qubit shows a correlation between the success
probability of finding the ground state, the single-qubit average and floppiness of the considered qubit.
For $\gamma=-1$, $A(t/t_a )=0$ and $B(t/t_a )=1$. Hence, the results for $\gamma$ close to -1 should not be taken into consideration.
Except for qubits $1$, $5$, $11$, the success probability for finding the ground state increases for a decreasing anneal offset (advanced annealing).
The floppiness of these qubits decreases monotonically for $\gamma<0$. When the floppiness is large,
the absolute value of the single-qubit average must be close to $0$. For qubits $1$, $5$, and $11$ the floppiness is small for $\gamma<0$.
For $\gamma>0$, the floppiness of these three qubits does not show any systematic behavior.

\subsection{Annealing offset applied to all qubits}

Based on the observations described in section~\ref{secOFFONE} and on the perturbative analysis~\cite{LANT17},
we developed strategies to enhance the probability of finding the ground state based on the value of
the floppiness and the absolute value of the single-qubit average. Both the floppiness and the absolute value
of the single-qubit average can be obtained from simulating the annealing process or
by performing the annealing process on a D-Wave machine.

We first consider the probability of floppiness $\mu_i$. We have tested various iterative methods based on the floppiness.
The best method we found so far is:
\begin{enumerate}
	\item Choose a static offset magnitude $\alpha=-0.02$.
	\item Initialize each anneal offset $\gamma_{i,0}$ to 0.
	\item For the iterations $k=1, \ldots, m$,
	\begin{enumerate}
			\item Perform a quantum annealing run and save $\mu_{i,k}$.
			\item Adjust each anneal offset, according to $\mu_{i,k}$: $\gamma_{i,k+1}=\gamma_{i,k}+\alpha \mu_{i,k}$.
	\end{enumerate}
\end{enumerate}
This method directly uses the information of $\mu_i$ from the previous run, and updates the anneal offset for each qubit accordingly.
We apply this method to three typical 2-SAT problems, namely to those with numbers $487$, $26$, and $301$
(see the Appendix for the explicit Hamiltonian $H_P$) having small, median,
and large minimal spectral gaps, respectively.

\begin{figure*}
\begin{center}
\includegraphics[width=0.48\hsize]{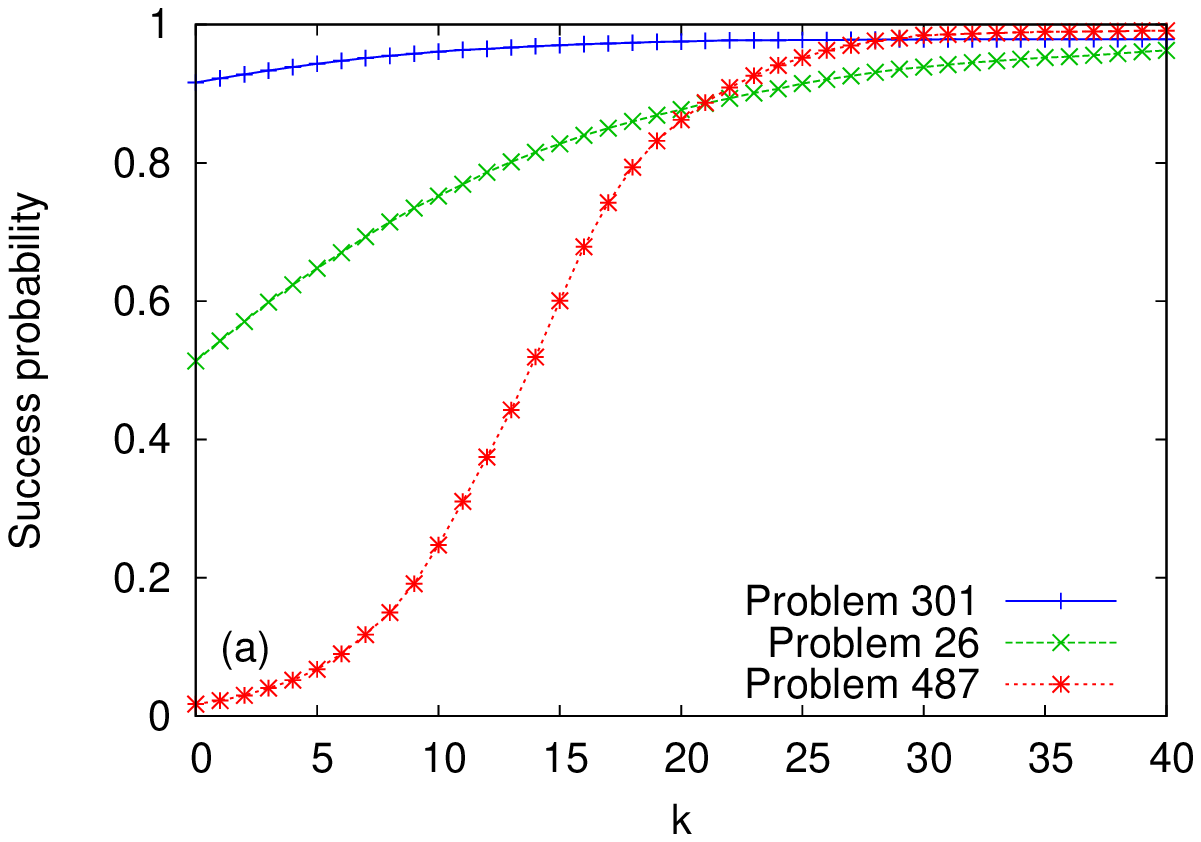}\includegraphics[width=0.48\hsize]{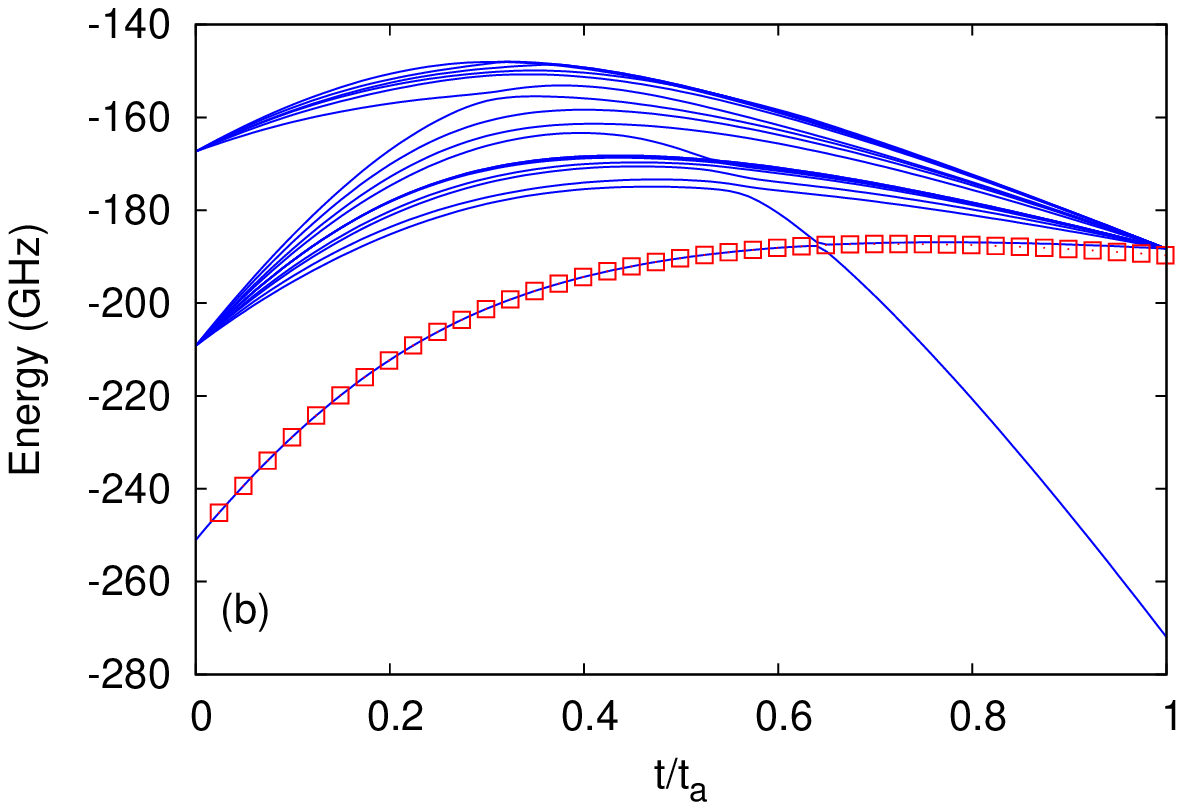}
\includegraphics[width=0.48\hsize]{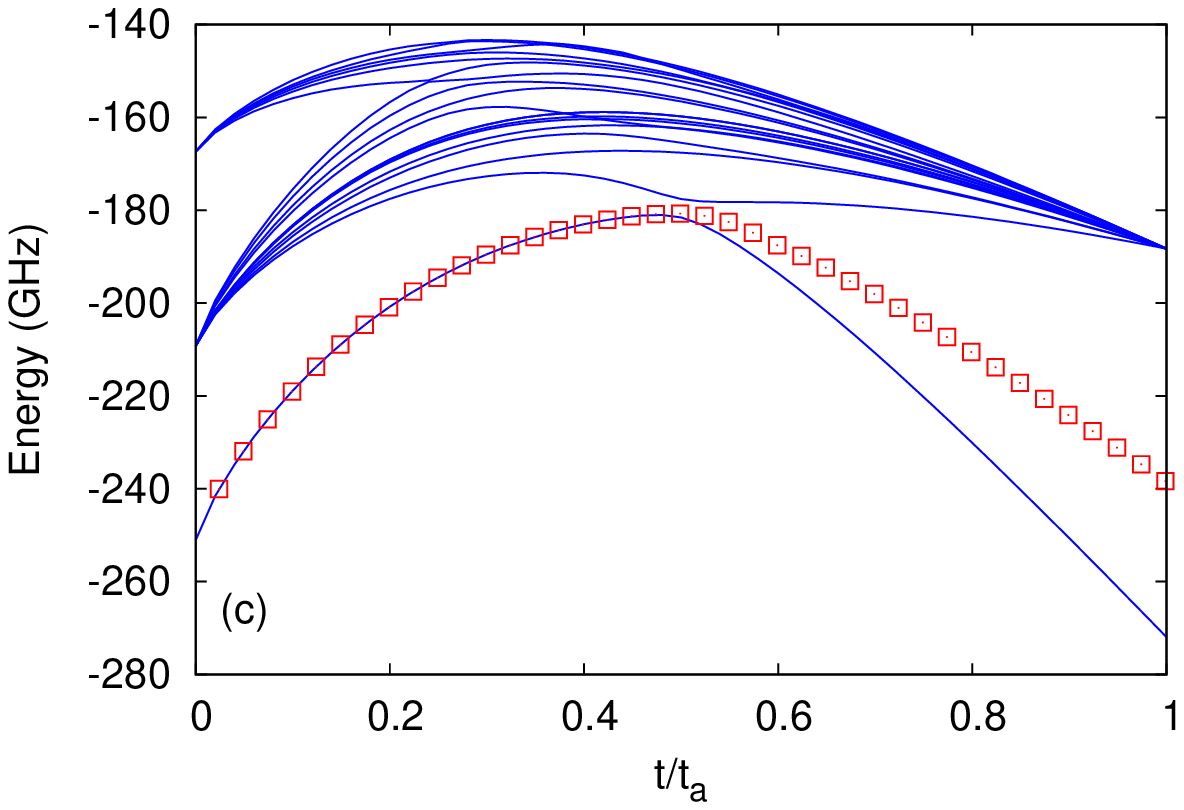}\includegraphics[width=0.48\hsize]{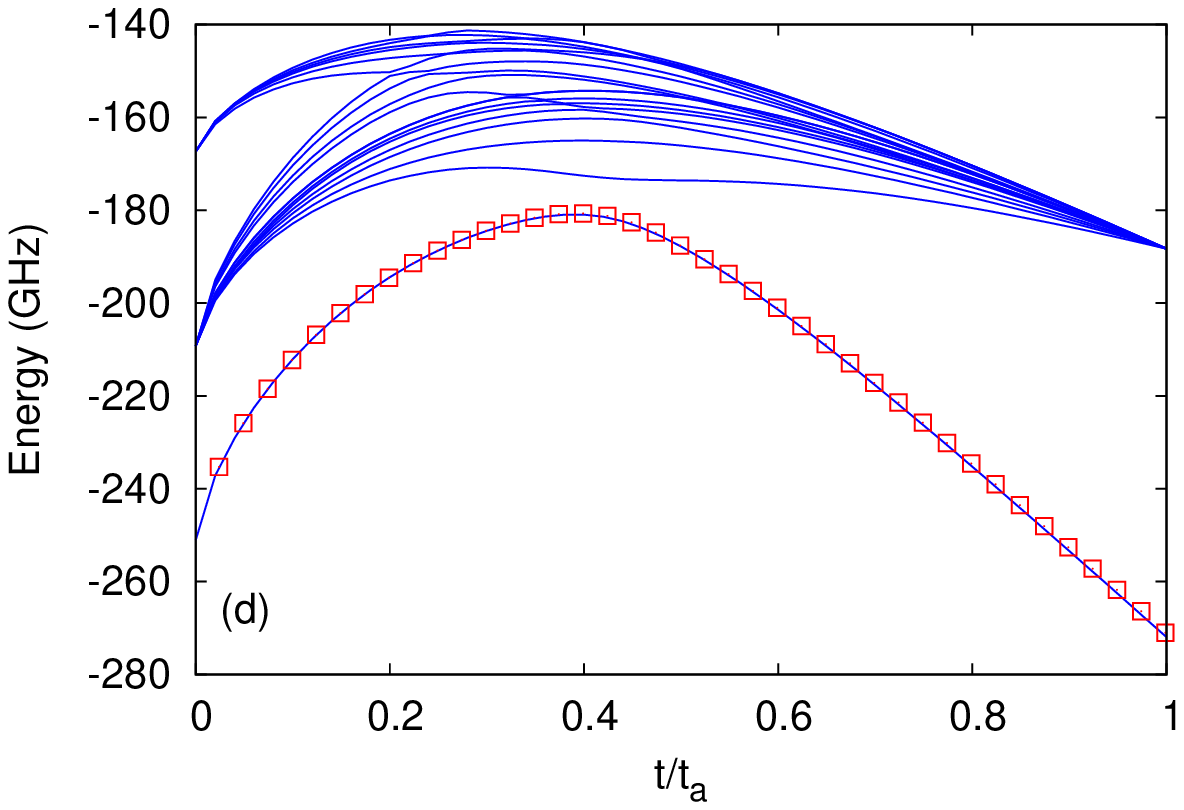}
\caption{%
(Color online) (a) Probability for finding the ground state by annealing for a time $t_a=5$ns as a function of the number
of iterations $k$ using an iterative method based on the floppiness for three different 2-SAT problems;
(b) energy spectrum of 2-SAT problem $487$  for a linear annealing process ($\Delta=0.84\;$GHz);
(c) annealing process with an anneal offset using $k=15$ iterations ($\Delta=3.97\;$GHz);
(d) annealing process with an anneal offset using $k=40$ iterations ($\Delta=8.36\;$GHz).
The red squares denote the average energy of the system during the annealing process.}
\label{fig3}
\end{center}
\end{figure*}

In Fig.~\ref{fig3}(a), we show for these three problems, the probability
for finding the ground state as a function of the number of iteration steps $k$ for the annealing time $t_a=5$ns.
For the three 2-SAT problems the success probability increases with the number of iterations, demonstrating
that the iterative method works. For the hardest problem (problem $487$), the change in the success probability
is of one order of magnitude. The big change in success probability can be understood from studying
the energy spectra and the average system energy for different numbers of iteration $k$ in the annealing process,
corresponding to different anneal offsets.
Figures 3(b-d) show the energy spectra for $k=1,15,$ and $40$, respectively.
With increasing iteration number $k$ the critical point shifts to the
left (decreasing $t/t_a$) and the minimal spectral gap increases. This indicates that the perturbative anticrossing
is eliminated. For the uniform linear annealing scheme, the average system energy follows the ground state energy
up to the critical point, then makes a Landau-Zener transition to end up close to the energy of the first excited
state of the problem Hamiltonian at the end of the annealing process. During the 15th iteration step,
the average system energy lies between the energy of the ground state and the first excited state after passing
the critical point. During the $40$th iteration step, the average system energy nicely follows the ground state
energy during the whole annealing process.

\begin{figure*}
\begin{center}
\includegraphics[width=0.48\hsize]{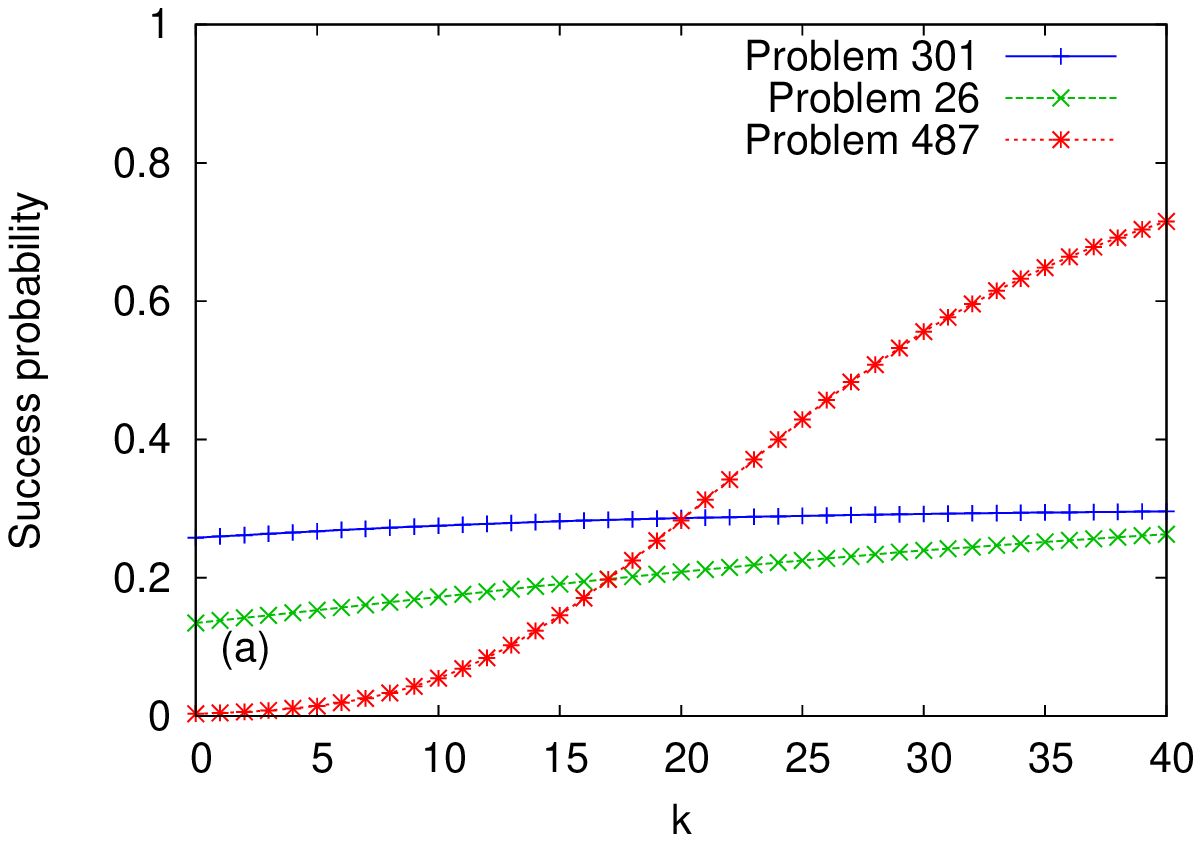}\includegraphics[width=0.48\hsize]{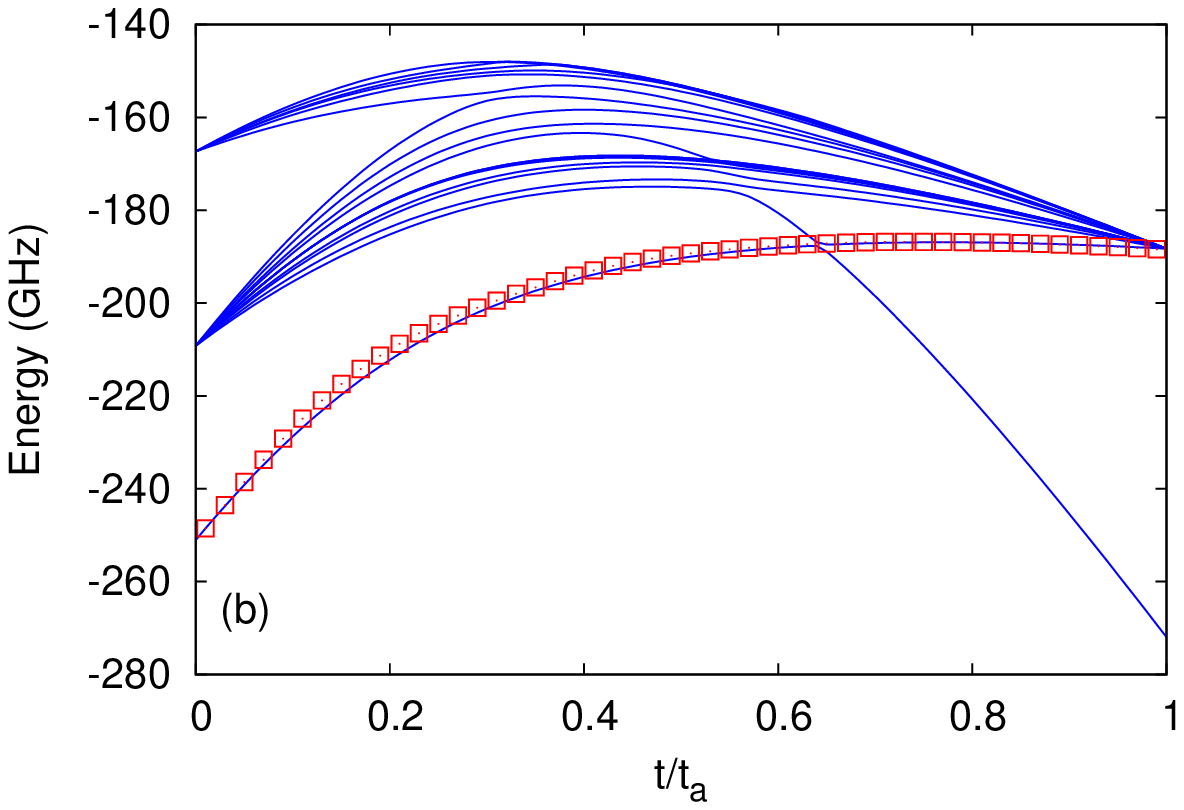}
\includegraphics[width=0.48\hsize]{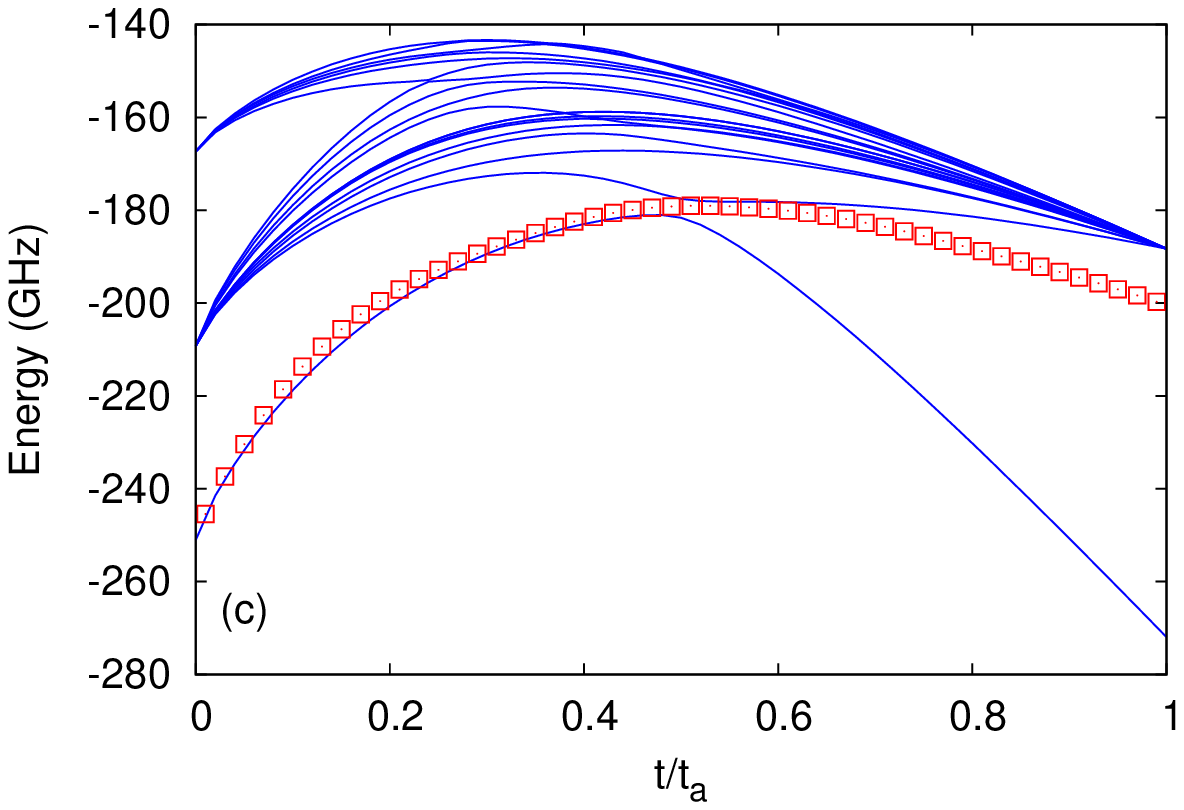}\includegraphics[width=0.48\hsize]{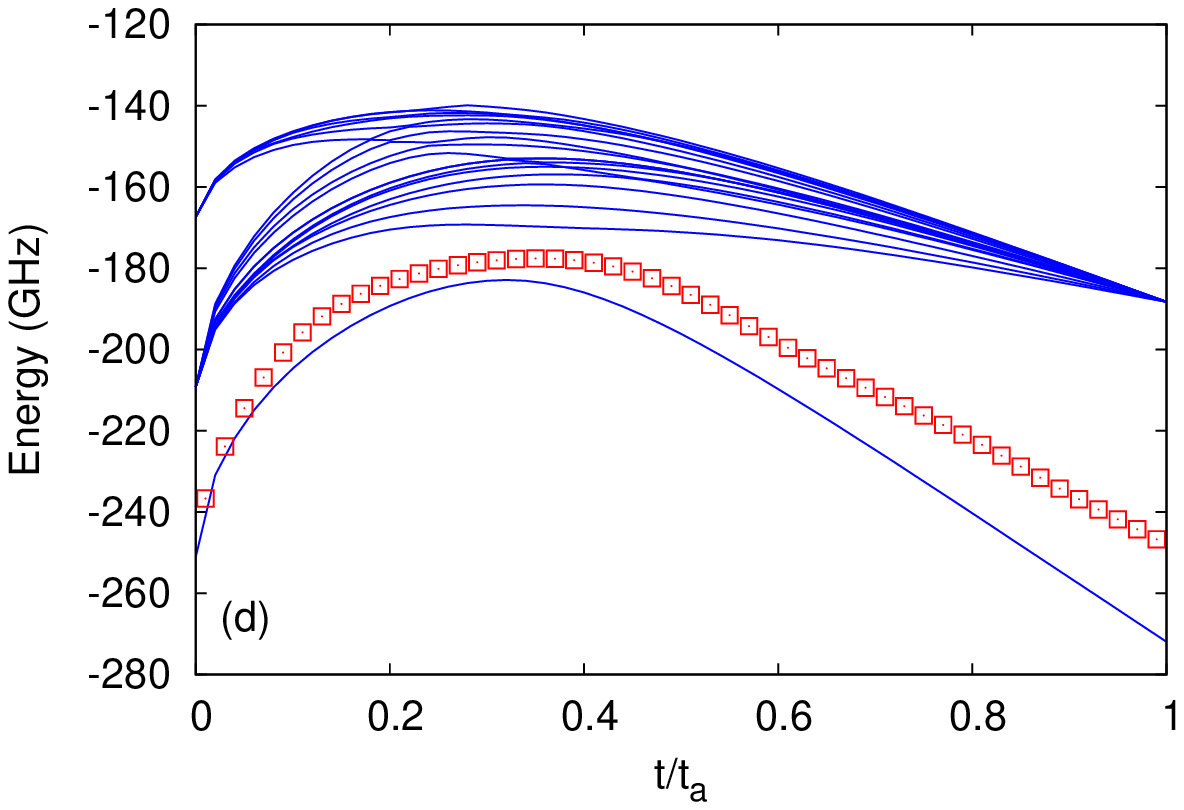}
\caption{%
(Color online) Same as Fig.~\ref{fig3} for $t_a=0.5$ns.
}
\label{fig4}
\end{center}
\end{figure*}

Results for a total annealing time $t_a=0.5$ns are shown in Fig.~\ref{fig4}. The results are similar to the ones shown
in Fig.~\ref{fig3}, but for none of the three problems the success probability reaches one for this short annealing time.
This agrees with the observation that even during the $40$th iteration step the average system energy does not follow the
ground state of the system during the annealing process. At the end of the annealing process the average system energy
lies in between the ground state energy and the energy of the first excited state of the problem Hamiltonian.

\begin{figure*}
\begin{center}
\includegraphics[width=0.48\hsize]{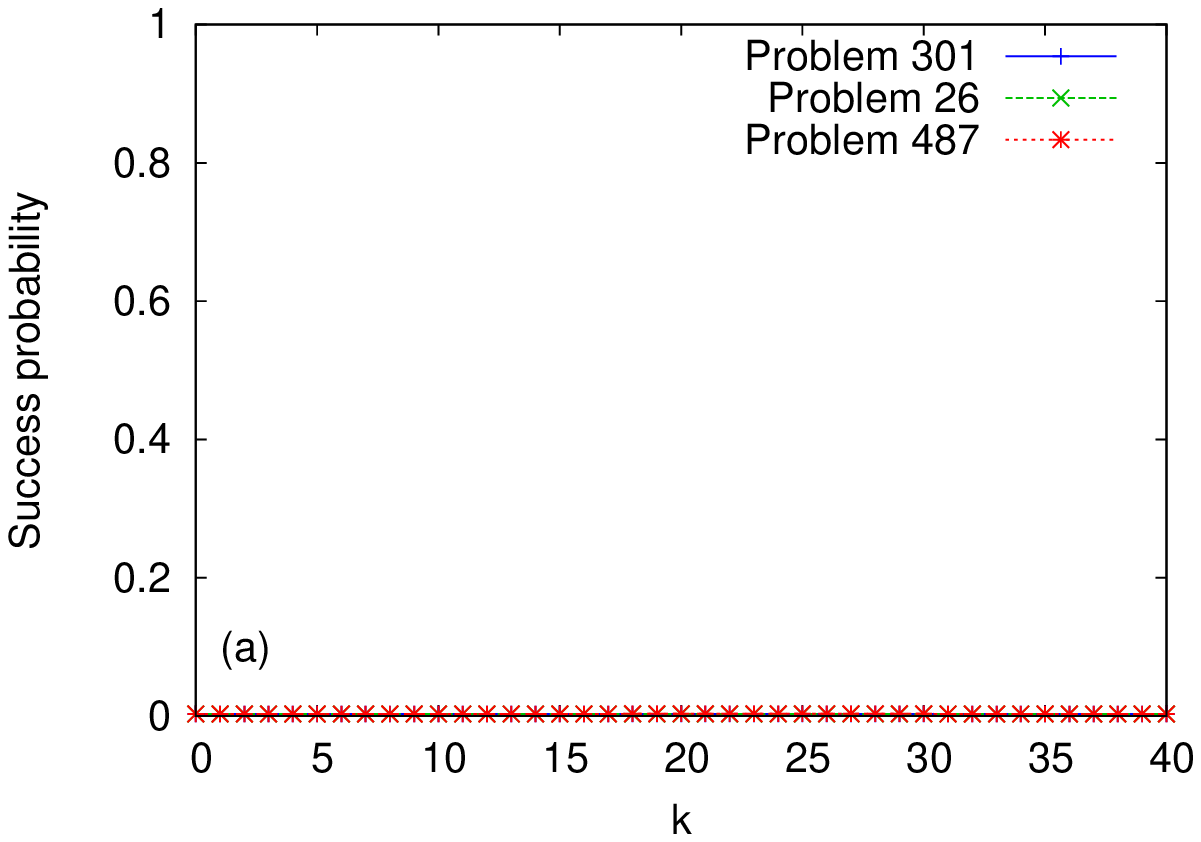}\includegraphics[width=0.48\hsize]{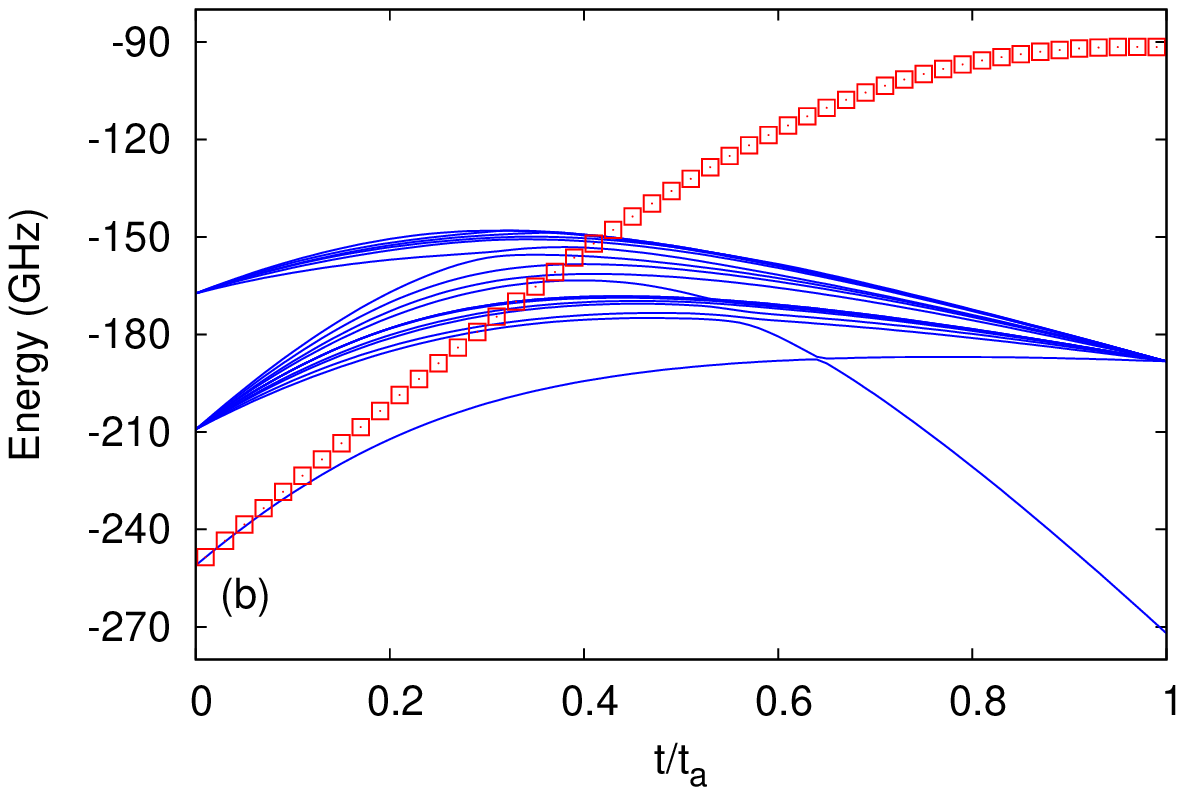}
\includegraphics[width=0.48\hsize]{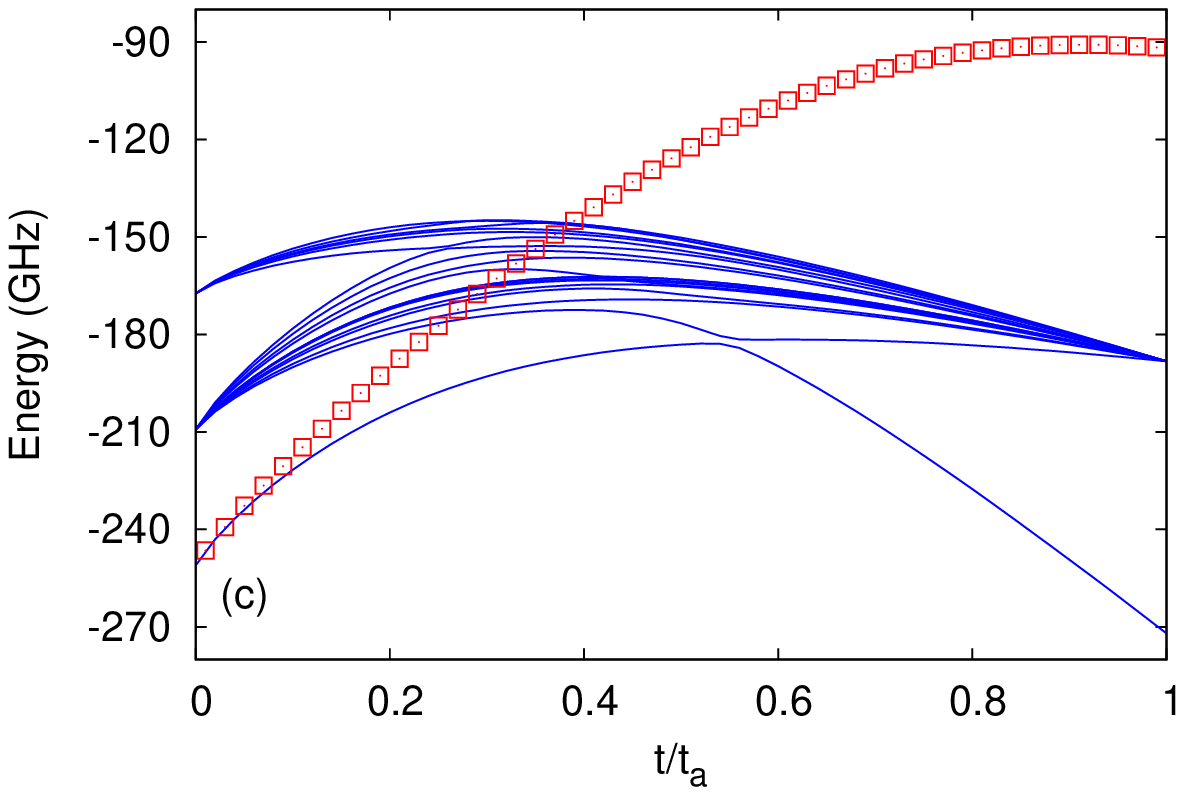}\includegraphics[width=0.48\hsize]{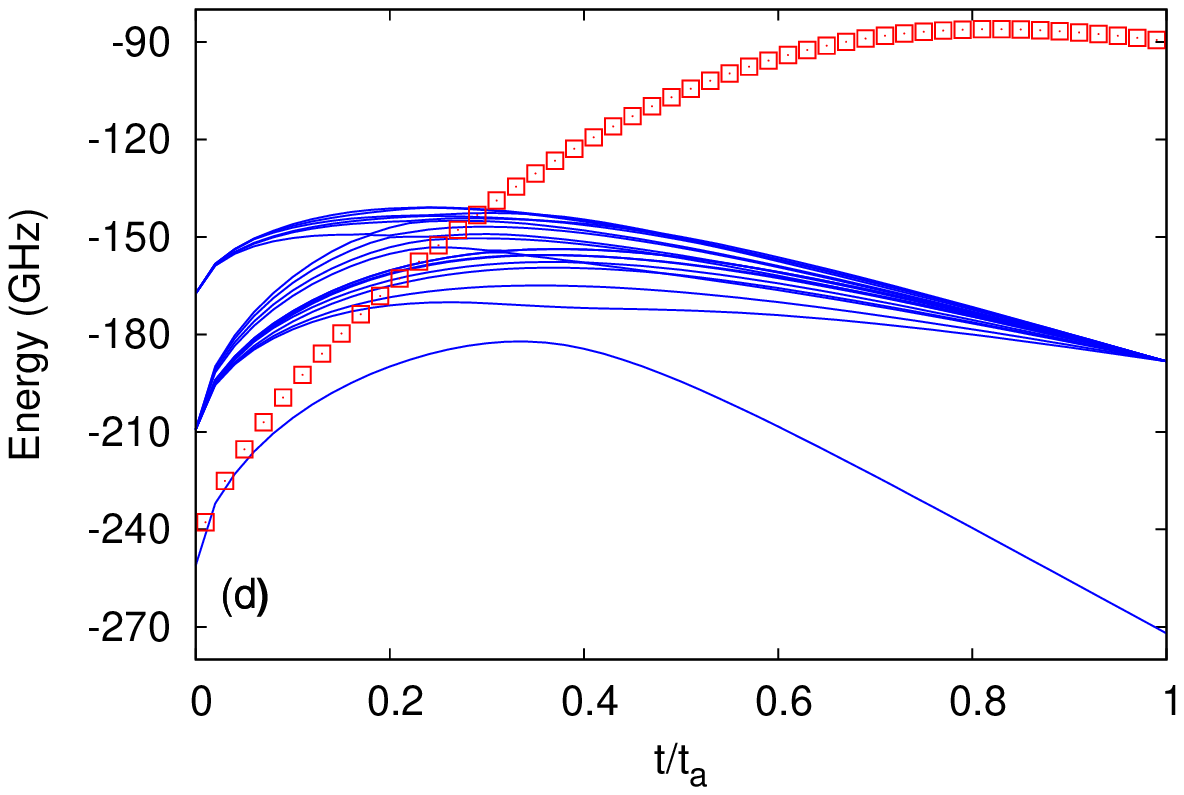}
\caption{%
(Color online) Same as Fig.~\ref{fig3} for $t_a=0.05$ns.
}
\label{fig5}
\end{center}
\end{figure*}

Results for an even shorter total annealing time $t_a=0.05$ns are shown in Fig.~\ref{fig5}.
These results are very different from the results shown in Figs.~\ref{fig3} and~\ref{fig4}.
First, for none of the three 2-SAT problems a reasonable success probability is obtained.
For all iteration numbers, the average energy of the system is rather high and corresponds to the one
of the higher excited states. Hence, the information obtained from the final sampling is far
from related to the first-excited state(s), a requirement for performing the perturbative analysis
and leads to a failure of the iterative method. Although the minimal spectral gap still opens
up with an increasing number of iterations, as seen from the figure, for such a fast annealing time
the average system energy is unable to follow the ground state of the system.

From Fig.~\ref{fig2}, we have seen that the behavior of the absolute value of the single-qubit average
shows some similarities with the success probability and the floppiness when varying the anneal offset for individual qubits.
For this reason, we might also use  $|\overline{\sigma}_i^z|$ to update the anneal offsets $\gamma_i$.
Some of the results are similar as those shown in Figs.~\ref{fig3},~\ref{fig4}, and~\ref{fig5} (results not shown).
However, some of them indicate worse performance compared to the performance using the floppiness of the qubits in
case the anneal offset is applied to all qubits. This means that the information coming from the absolute value
of the single-qubit average cannot be used reliably to define an iterative annealing process with an anneal offset.
Therefore, we do not show any results based on the absolute value of the single-qubit average.

The above results strongly suggest that only information obtained from the first excited states is useful
to improve the outcome of the annealing process by an anneal offset. If instead of the global minimum,
only a near-optimal solution is desired, we already obtain this near-optimal
solution from the first excited states. In terms of energy, the near-optimal solution is very close to the optimal
solution and one could try to use a stochastic approximation algorithm, such as the Kiefer-Wolfowitz algorithm,
to find this optimal solution starting from the near-optimal solution. We use as the cost function for updating the anneal offsets,
the average energy $\overline{E}_f$, obtained from the final events. The method works as follows
\begin{enumerate}
	\item Initialize each anneal offset $\gamma_{i,0}$ to 0.
	\item For iteration $k=1,\ldots, m$,
\begin{enumerate}
	\item For each spin, perform two quantum annealing runs with $\gamma_{i,k}=\gamma_{i,k}+c_k$ and $\gamma_{i,k}=\gamma_{i,k}-c_k$,
	and save the corresponding average energy as $\overline{E}_f(\gamma_{i,k}+c_k)$ and $\overline{E}_f(\gamma_{i,k}-c_k)$.
	\item Adjust each anneal offset as follows
	\begin{equation}
	\gamma_{i,k}=\gamma_{i,k}+\alpha_k \left ( \frac{\overline{E}_f(\gamma_{i,k}+c_k)-\overline{E}_f(\gamma_{i,k}-c_k)}{2c_k}  \right ).
	\end{equation}
\end{enumerate}
\end{enumerate}
The parameters $\alpha_k$ and $c_k$ are taken to be $k^{-1}$ and $k^{-1/3}$, respectively.
From the update formula for the anneal offset it can be seen that the Kiefer-Wolfowitz algorithm can be considered as a
gradient-like method with the gradient generated by a finite difference scheme.

\begin{figure*}
\begin{center}
\includegraphics[width=0.32\hsize]{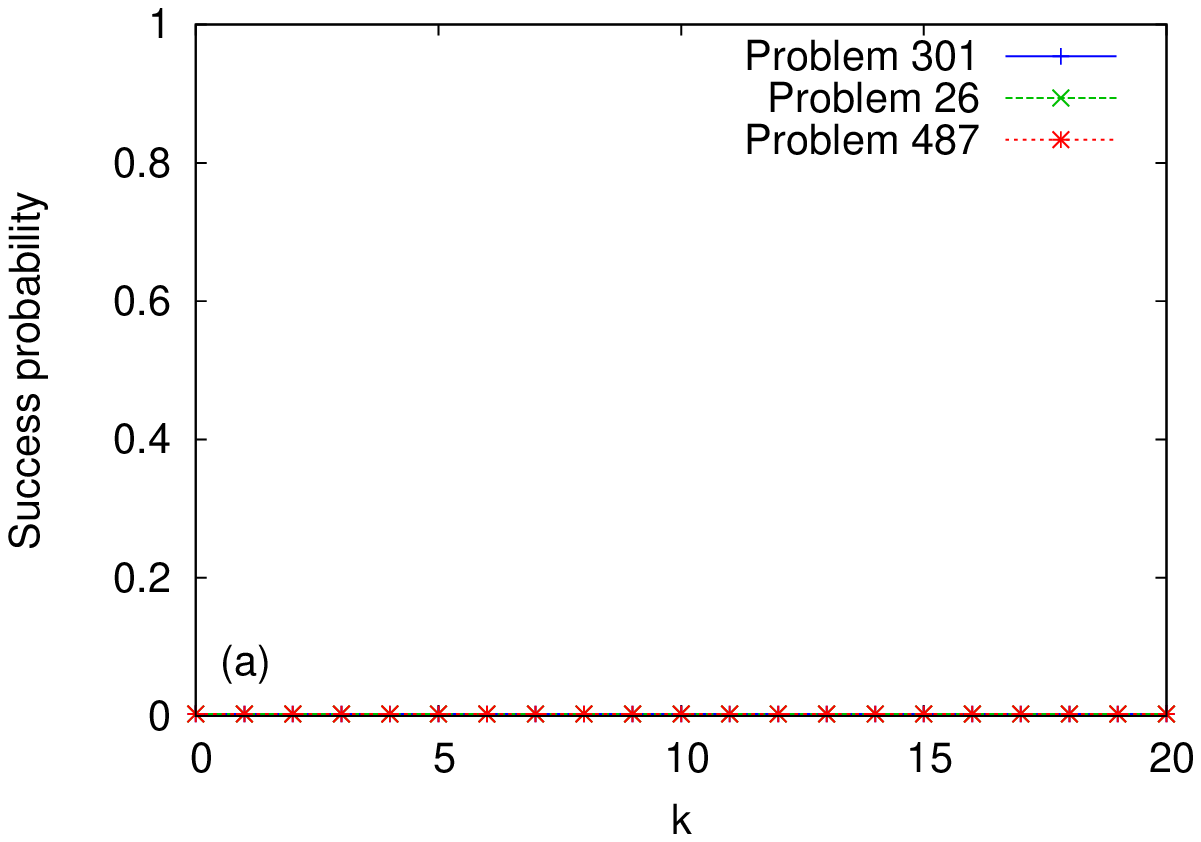}
\includegraphics[width=0.32\hsize]{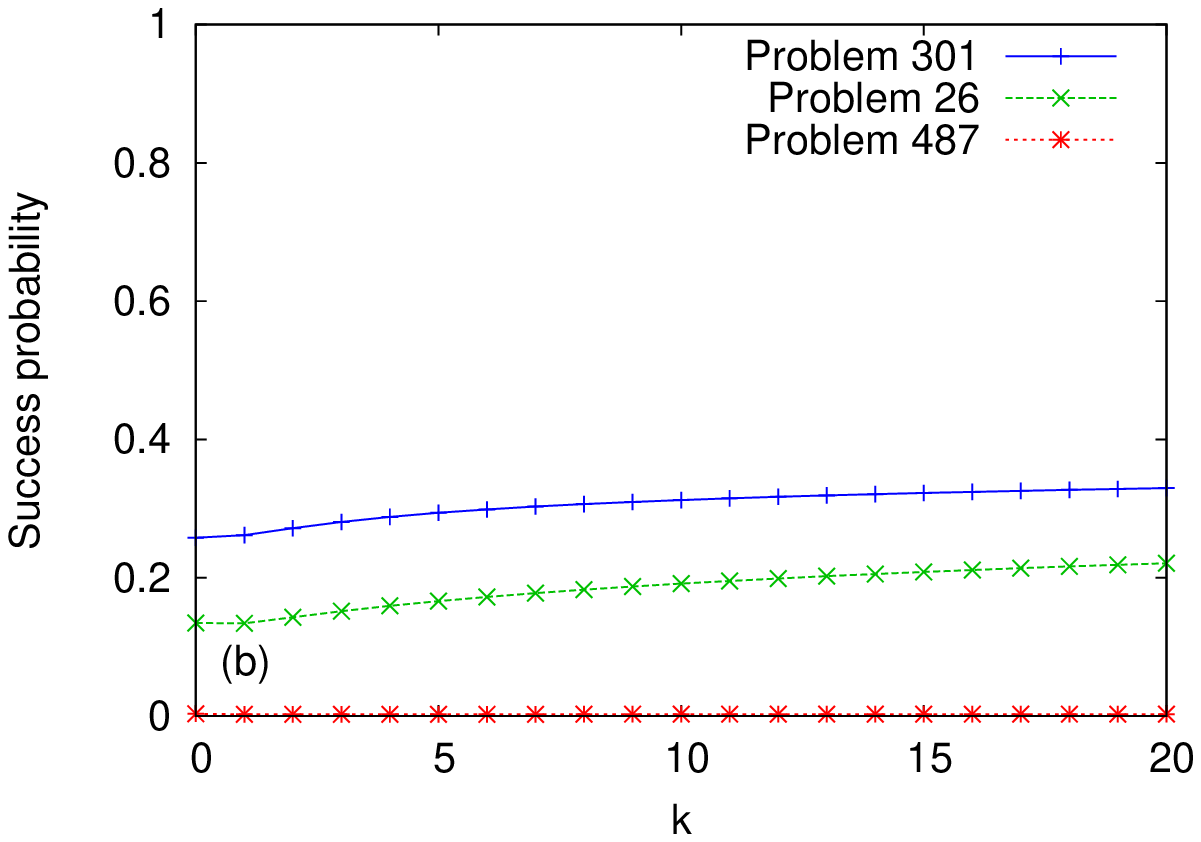}
\includegraphics[width=0.32\hsize]{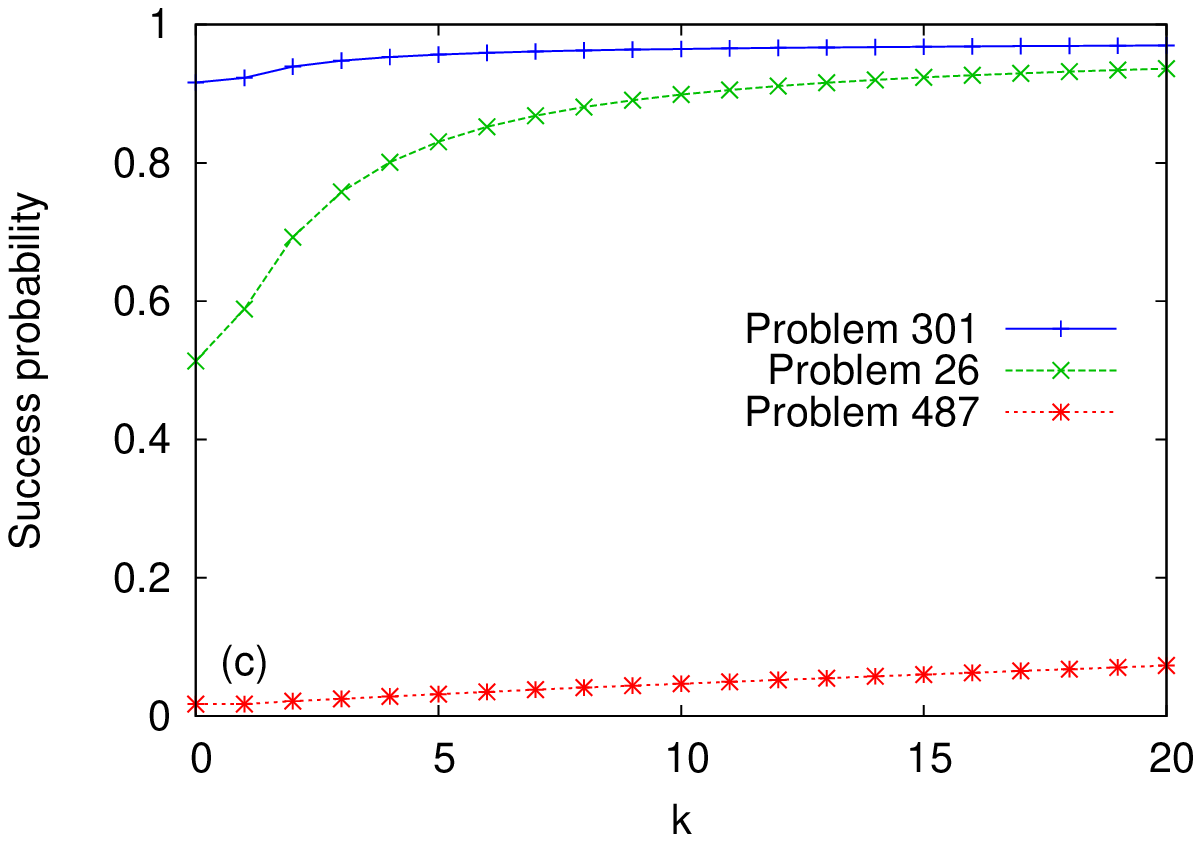}
\caption{%
(Color online) Probability for finding the ground state of three different 2-SAT problems as a function of the number
of iterations $k$ using an iterative method based on the Kiefer-Wolfowitz algorithm for different annealing times:
(a) $t_a=0.05$ns; (b) $t_a=0.5$ns; and (c) $t_a=5$ns.
}
\label{fig6}
\end{center}
\end{figure*}

The results of applying this iterative method for solving the three 2-SAT problems are shown in Fig.~\ref{fig6}.
At first sight, the performance of the Kiefer-Wolfowitz algorithm competes with the iterative method
based on the floppiness of the qubits (see Figs.~\ref{fig3}, \ref{fig4}, and \ref{fig5}).
However, the method based on the Kiefer-Wolfowitz algorithm is computational
intensive because within each iteration, the number of quantum annealing runs to be performed is proportional
to the size of the spin system. An interesting case is 2-SAT problem $487$, the hardest instance. If, for this problem,
the anneal offset is tuned by the iterative method based on the spin floppiness, then the probability to find the ground
state shows a very big increase as a function of the number of iterations (see Fig.~\ref{fig3}(d))
compared to the increase obtained for the other two 2-SAT instances. If, for problem $487$, the anneal offset is
obtained by the iterative method based on the Kiefer-Wolfowitz algorithm, then the success probability does not show this big
increase as a function of the number of iterations. An explanation for this observation could be that the parameters
$\alpha_k$ and $c_k$ are not well chosen because the energy differences become rather small for this hard problem.
Therefore, without a proper procedure to choose the parameters $\alpha_k$ and $c_k$, the algorithm may require much more iteration steps to
converge to a (near) optimal solution. For a long annealing time ($t_a=5$ns) the final state has a large
overlap with the first excited state. This means that at the end of the annealing process the system is in a state close
to the global minimum of the (classical) problem Hamiltonian and then the gradient-like method works well.
If the final system state is far away from the global minimum, as is the case for the short annealing times,
then this algorithm will not work because then it is easy to get stuck in local minima.

From the analysis of the simulation results, it is clear that the best strategy to enhance the performance of the quantum
annealing process is to first anneal with longer annealing times.
If annealing longer (in practice one can only anneal for a given fixed time)
does not help, which could be due to various effects such as temperature effects for example, then one could consider
to perform quantum annealing with an adaptive anneal offset determined by an iterative method.
However, from the investigation of the average system energy as a function of the number of iterations
it is also clear that this approach only works if at the start of the iterative process the average system energy lies
in between the ground state energy and the energy of the first exited state. In other words, the average system energy
should fall in the Landau-Zener regime. This constraint is also an assumption in the degeneracy calculation in Ref.~\cite{LANT17}.

\begin{figure*}
\begin{center}
\includegraphics[width=0.32\hsize]{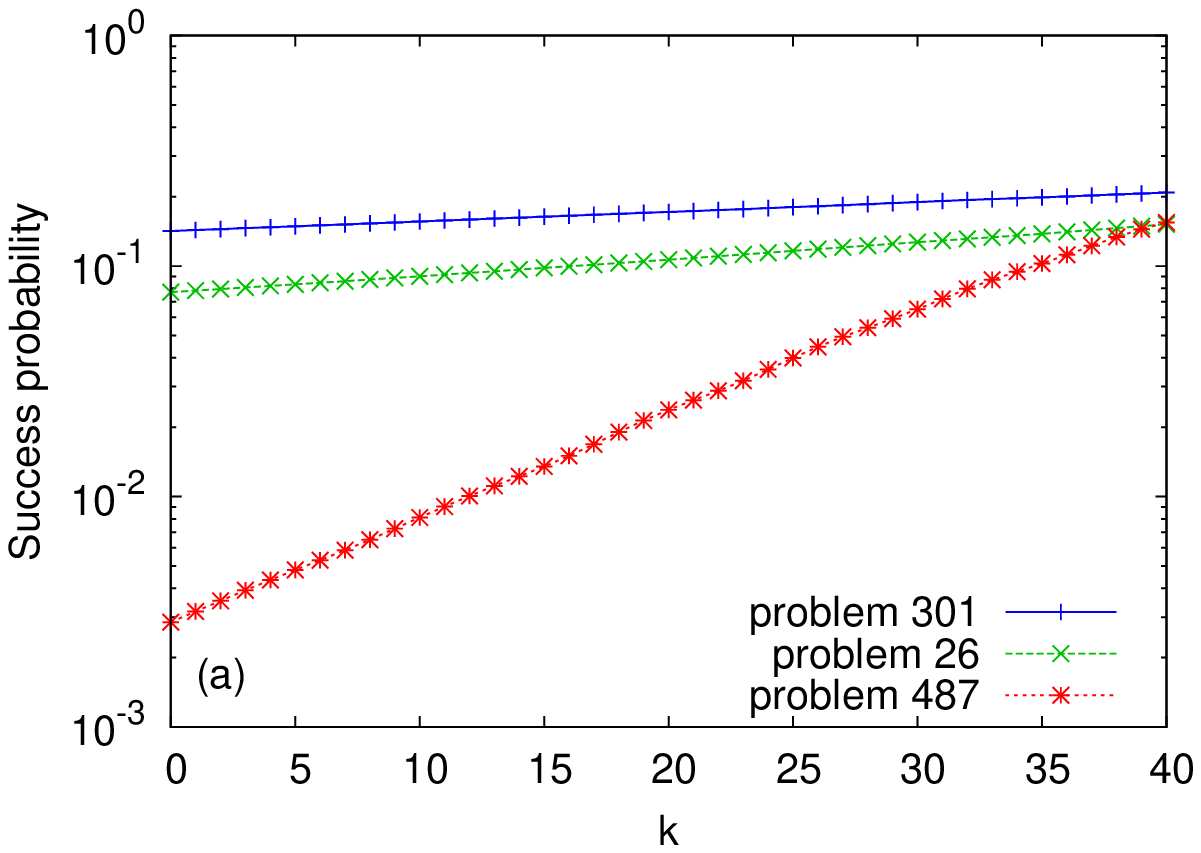}
\includegraphics[width=0.32\hsize]{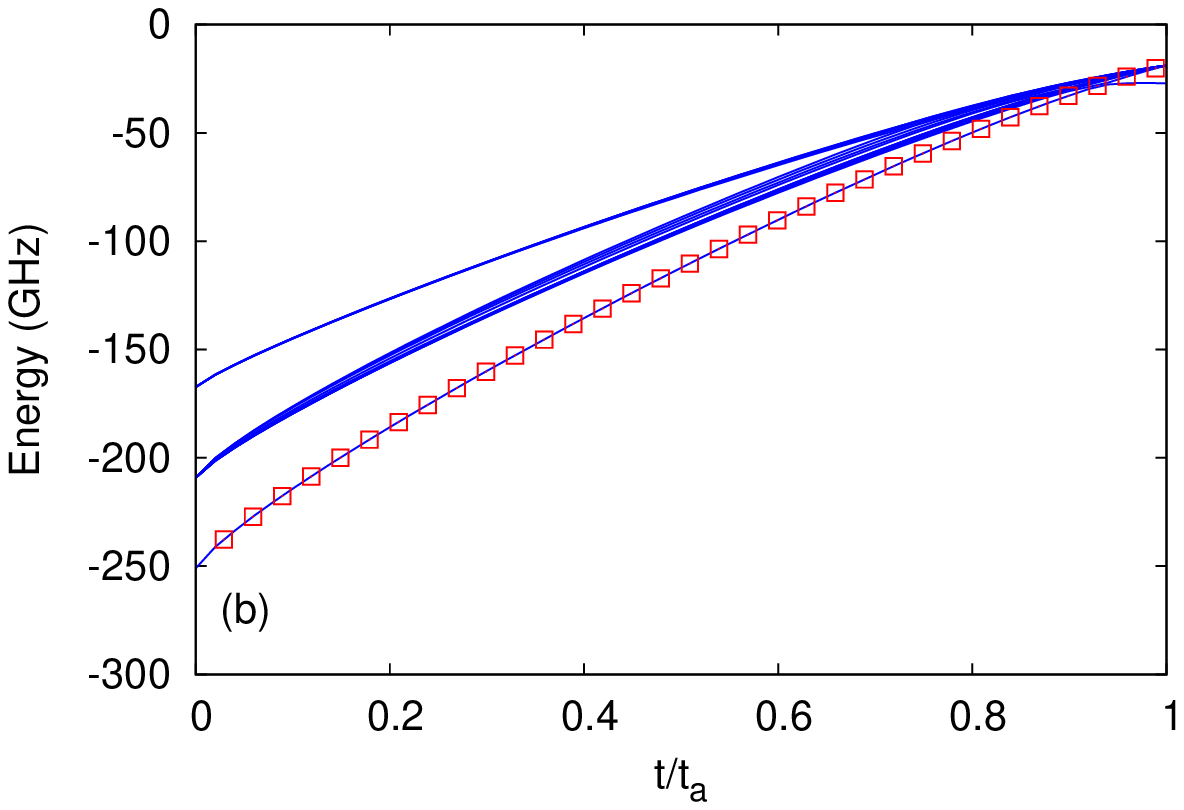}
\includegraphics[width=0.32\hsize]{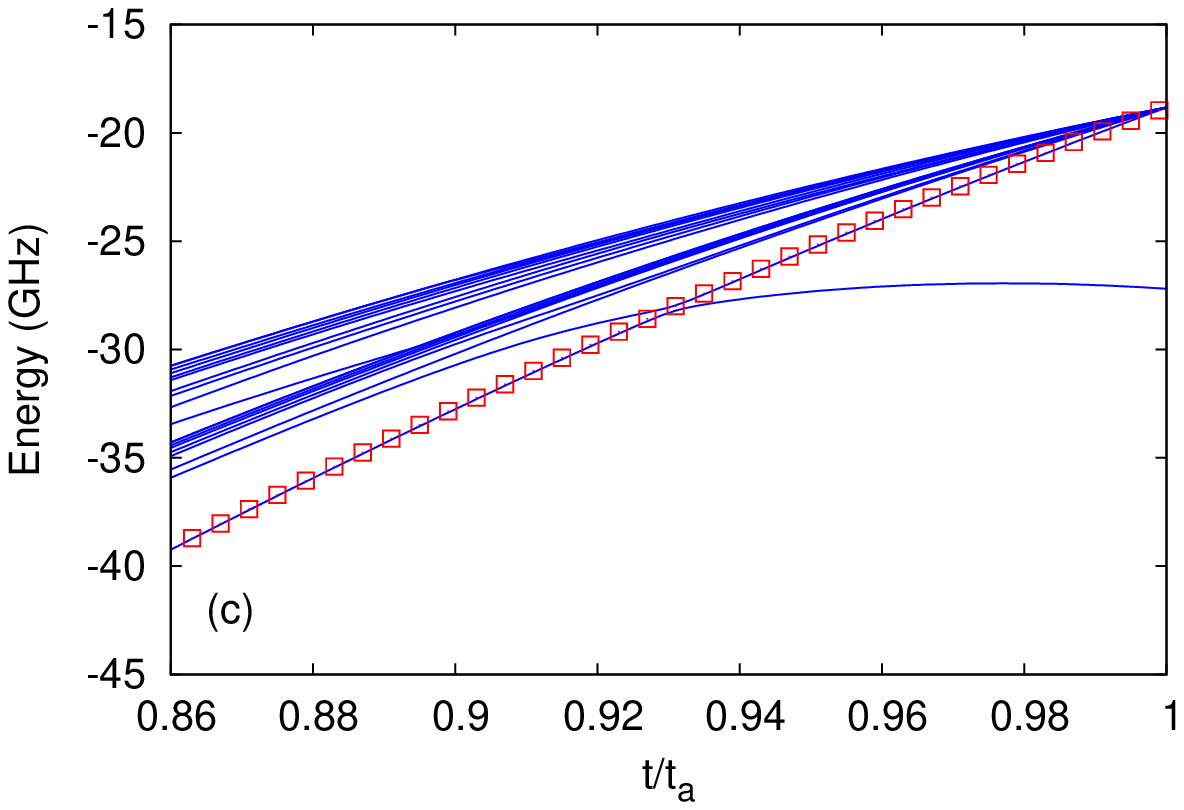}
\includegraphics[width=0.32\hsize]{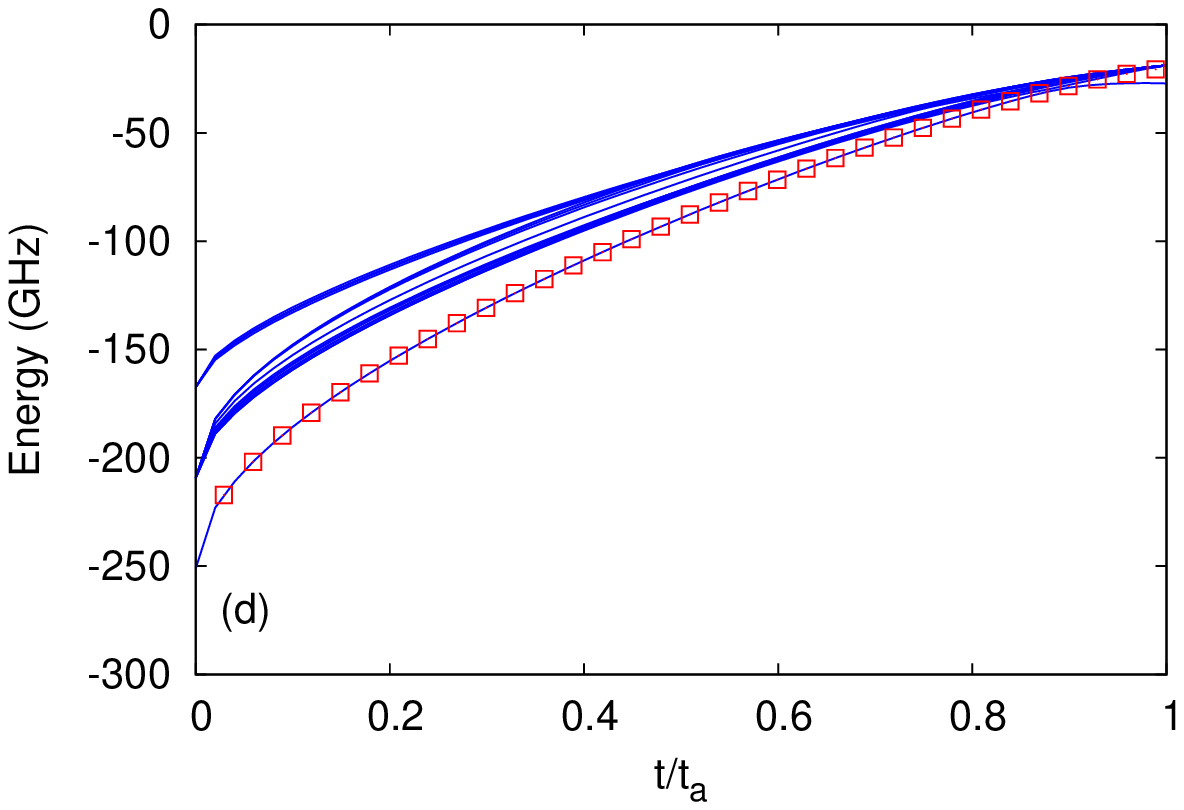}
\includegraphics[width=0.32\hsize]{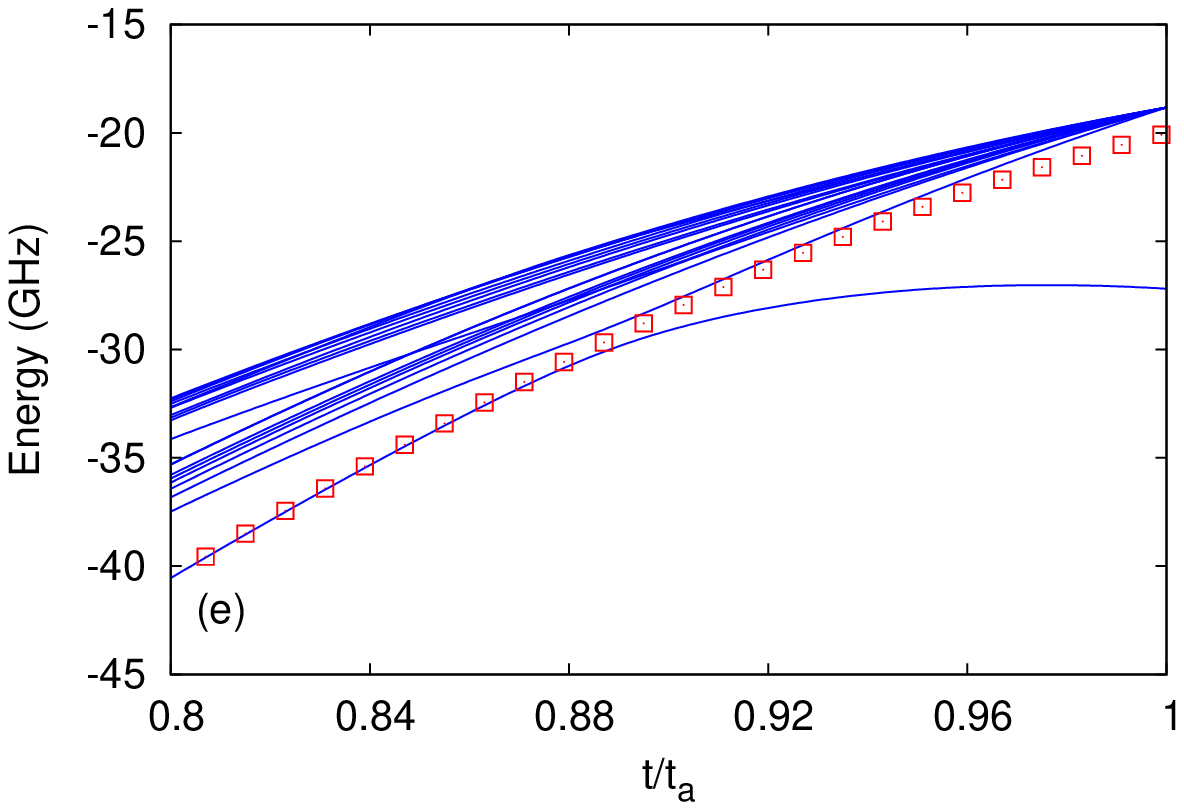}
\caption{%
(Color online) (a) Probability for finding the ground state by annealing for a time $t_a=5$ns as a function of the number of
iterations $k$ using an iterative method based on the floppiness for three 2-SAT problems with $C=0.1$;
(b,c) energy spectrum of the 2-SAT problem $487$ for an annealing process with an anneal offset for $k=15$ iterations;
(d,e) energy spectrum of the 2-SAT problem $487$ for an annealing process with an anneal offset for $k=40$ iterations.
The red squares denote the average energy of the system during the annealing process.
}

\label{fig7}
\end{center}
\end{figure*}

\begin{figure*}
\begin{center}
\includegraphics[width=0.32\hsize]{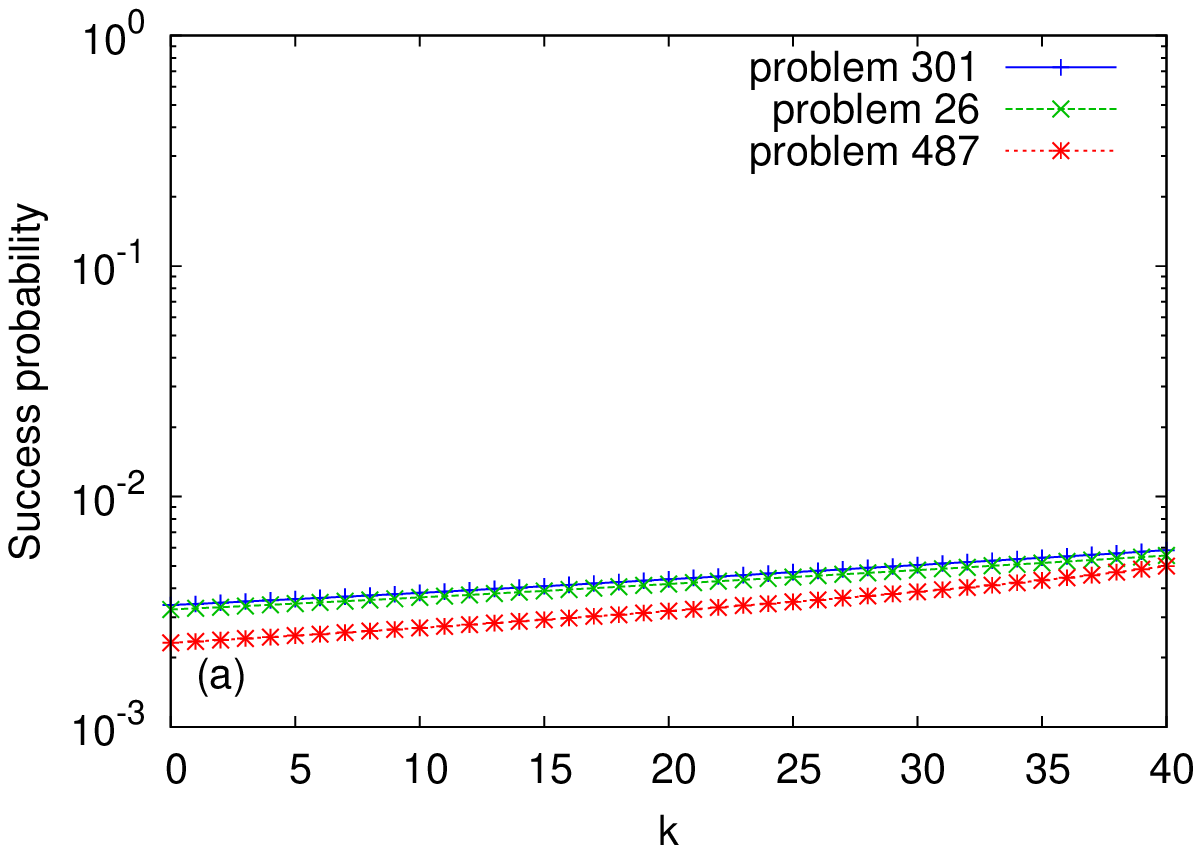}
\includegraphics[width=0.32\hsize]{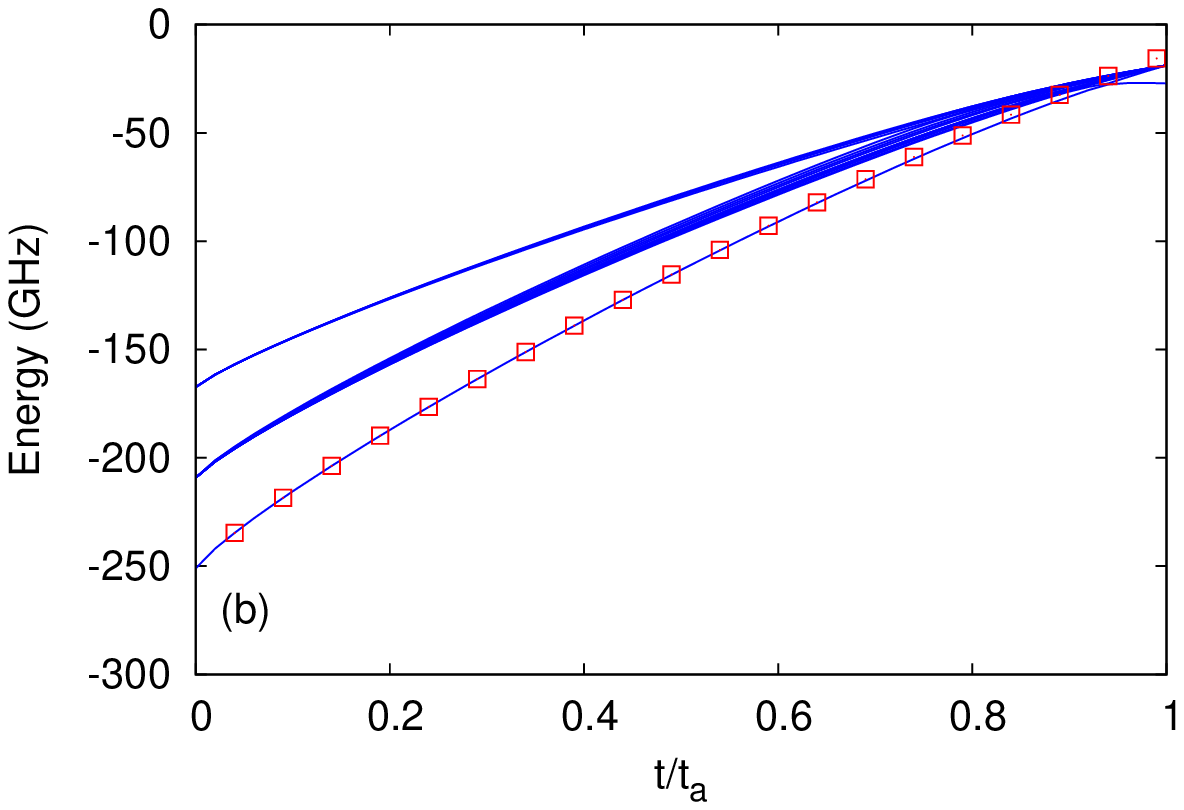}
\includegraphics[width=0.32\hsize]{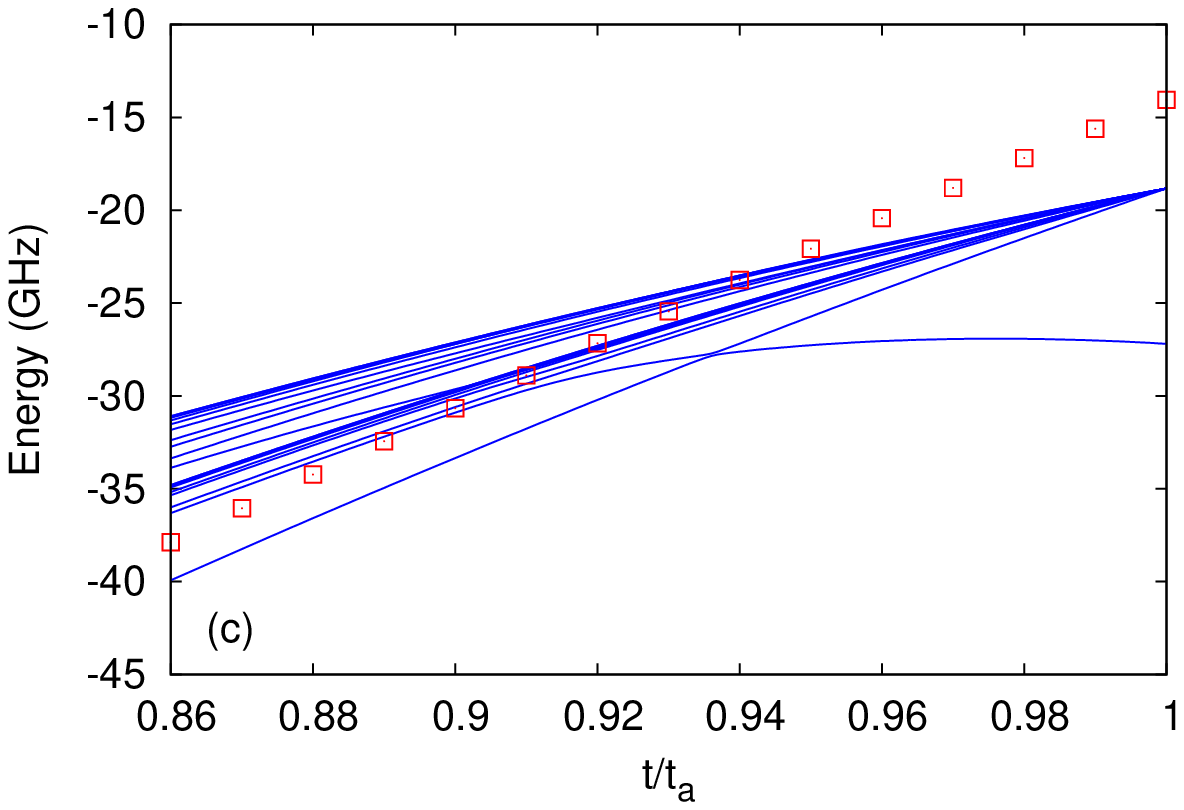}
\includegraphics[width=0.32\hsize]{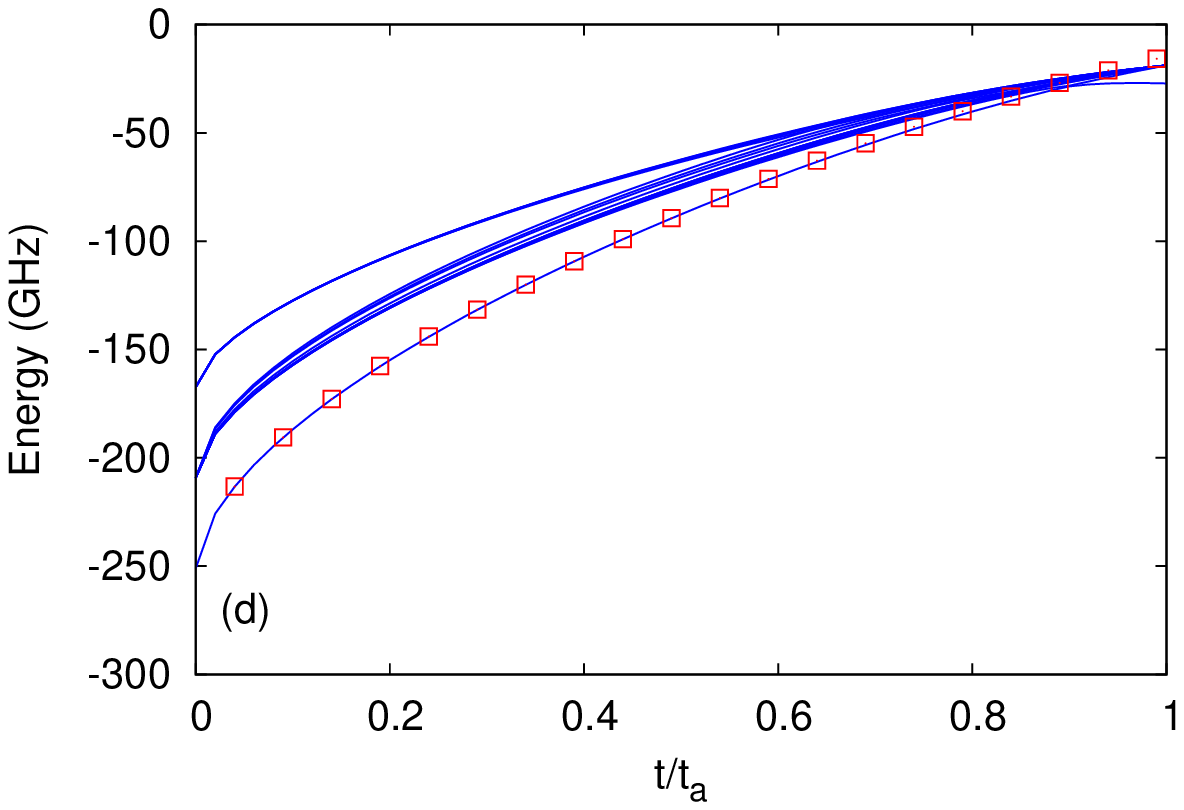}
\includegraphics[width=0.32\hsize]{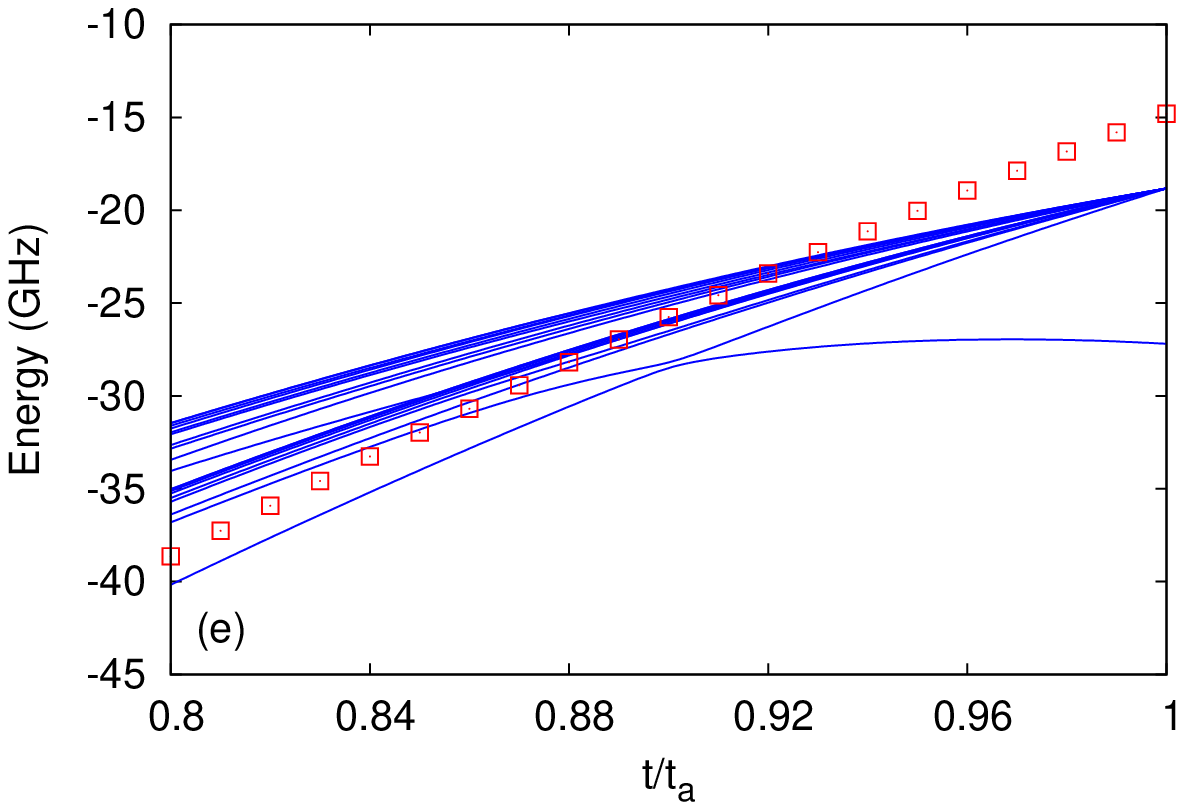}
\caption{%
(Color online) Same as Fig.~\ref{fig7} for $t_a=0.5$ns.
}
\label{fig8}
\end{center}
\end{figure*}

For testing the performance of the iterative method based on the spin floppiness on even harder problems,
we reduce the minimal spectral gap of the 2-SAT problems by introducing a factor $C$
\begin{equation}
H=A(t/t_a) H_I+B(t/t_a )(C H_P).
\end{equation}
By assigning to $C$ a value smaller than one, the energy gap between the ground state and the first excited state shrinks.
We consider $C=0.1$.
Figures~\ref{fig7} and~\ref{fig8} depict the results for annealing times $t_a=5$ns and $t_a=0.5$ns, respectively.
Comparing the energy spectra with those presented in Figs.~\ref{fig3} and~\ref{fig4} shows that the critical point moves
to the right and that the minimum spectral gap is smaller. This means that the problem indeed became harder and that this
situation corresponds even better to a perturbative anticrossing. Also for this case it is clear that the iterative approach
only works if at the start of the iterative process the average system energy falls in the Landau-Zener regime.

\section{Conclusions}\label{secCONCLUSION}

The quantum annealing method with an adaptive anneal offset is practical method for enhancing the performance
of the annealing method for systems for which perturbative anticrossings dominate the slow-down mechanism of quantum annealing.
For such systems, like for example the special 2-SAT problems that we considered, the first excited state is largely degenerate.
Adjusting the anneal offset for qubits with a high floppiness can enlarge the minimal spectral
gap by one or two orders of magnitude and therefore enhance the performance of the quantum annealing method.

Under certain circumstances, as for example when a system is coupled to an environment,
simply increasing the annealing time will not tremendously improve the probability for finding the ground state.
In such a case, an iterative quantum annealing method based on the floppiness of the qubits could
be considered to drive the system to the ground state. For comparison, we also implemented
the Kiefer-Wolfowitz algorithm. We have demonstrated that it is a stable method with the drawback
that the required number of annealing runs is proportional to the system size.

Note that in this work, we did not investigate the scaling behavior of the iterative quantum annealing
method based on the floppiness of the qubits. This would require a systematic problem-size-dependent investigation.
In this respect, it is worth noting that for a particular
mean-field-type model, Susa et al. showed that a certain type of inhomogeneous driving
of the transverse field erases first-order quantum phase transitions,
and brings an exponential speedup for quantum annealing~\cite{SUSA18}.

\section*{Acknowledgement}
Research was partially funded by Volkswagen Group, department Group IT.

\appendix
\section*{Appendix}
Here we present the detail of the problem Hamiltonian investigated in the main text.
The parameters $h_i^z$ and $J_{ij}^z$ in $H_P$ (see Eq.~(\ref{problemH})) are listed in Tables~\ref{table1} and ~\ref{table2}.

\begin{table*}[t]
\caption{The parameters $h_i^z$ in $H_P$ (see Eq.~(\ref{problemH})) for 2-SAT problems with $12$ Boolean variables
with numbers $487$, $26$, and $301$.
}
\begin{center}
\begin{tabular}{c|c|c|c}
\hline
        &   Problem $487$  & Problem $26$  & Problem $301$   \\
\hline
    $i$ &{$h_i^z$} &{$h_i^z$} &{$h_i^z$}\\
\hline
    1 &  -2 &  0 &  0 \\
    2 &  -1 & -1 &  0 \\
    3 &		1 &  0 &  0 \\
    4 &  -1 & -1 &  0 \\
    5 &  -2 &  1 &  1 \\
    6 &   1 & -1 &  0 \\
    7 &  -1 &  0 &  1 \\
    8 &   1 &  1 &  0 \\
    9 &   1 &  0 &  1 \\
    10 &  1 & -1 &  1 \\
    11 & -2 &  0 &  0 \\
    12 &  0 &  0 &  0 \\
\hline
\end{tabular}
\label{table1}
\end{center}
\end{table*}

\begin{table*}[t]
\caption{The parameters $J_{ij}^z$ in $H_P$ (see Eq.~(\ref{problemH})) for 2-SAT problems with $12$ Boolean variables
with numbers $487$, $26$, and $301$.
}
\begin{center}
\begin{tabular}{cc|c|cc|c|cc|c}
\hline
    \multicolumn{3}{c|}{Problem $487$} & \multicolumn{3}{c|}{Problem $26$} & \multicolumn{3}{c}{Problem $301$}   \\
\hline
   $i$ & $j$ & $J_{ij}^z$     & $i$ & $j$ & $J_{ij}^z$ & $i$ & $j$ & $J_{ij}^z$   \\
\hline
   1 & 4 & -1 &   1 & 9 & 1 & 		1 & 2 & -1 \\
   1 & 6 &  1 &   1 & 11 & -1 &		1 & 3 & 1 \\
   1 & 8 &  1 &   2 & 11 & 1 &		2 & 4 & -1 \\
   1 & 11 & 1 &   3 & 4 & 1 &			3 & 8 & 1 \\
   2 & 6 &  0 &   3 & 6 & 1 & 		4 & 12 & 1 \\
   2 & 11 & -1 &  4 & 8 & 1 &			5 & 12 & 1 \\
   3 & 5 &  1 &   4 & 10 & -1 &		6 & 7 & 1 \\
   5 & 7 & -1 &   5 & 12 & -1 &		6 & 10 & -1 \\
   5 & 11 & 1 &   6 & 7 & -1 &		8 & 11 & -1 \\
   5 & 12 & 1 &   6 & 12 & -1 &		9 & 10 & -1 \\
   9 & 11 & 1 &   7 & 9 &  -1 &		10 & 11 & 1 \\
   10& 12 & 1 &   8 & 12 & 0 &   	11 & 12 & 0 \\
   1 & 2 &  0 &   1 & 2 & 0 &     1 & 4 & 0 \\
\hline
\end{tabular}
\label{table2}
\end{center}
\end{table*}
\bibliography{TwoSAT}   

\end{document}